\documentclass[longauth,colorlinks=true,linkcolor=black,citecolor=blue,urlcolor=blue]{aa}
\usepackage{txfonts}
\usepackage{graphicx} 
\usepackage{multirow}
\usepackage{xcolor}
\usepackage{booktabs}
\usepackage{orcidlink}
\usepackage{color}
\usepackage{placeins}
\usepackage[version=4]{mhchem}
\usepackage{ulem}
\usepackage[autostyle]{csquotes}
\usepackage{amsmath}
\usepackage{amssymb}
\begin{document} 

   \title{MINDS. Hydrocarbons detected by JWST/MIRI in the inner disk of Sz28 consistent with a high C/O gas-phase chemistry}
   \titlerunning{Hydrocarbons detected by JWST/MIRI in inner disk of Sz28 consistent with high a C/O gas-phase chemistry}

\author{{Jayatee Kanwar} \orcidlink{0000-0003-0386-2178}
          \inst{1,2,3}
           \and
          Inga Kamp \orcidlink{0000-0001-7455-5349} \inst{1} \and
          Hyerin Jang \orcidlink{0000-0002-6592-690X} \inst{4} \and
          L.B.F.M. Waters \orcidlink{0000-0002-5462-9387} \inst{4,5} \and 
          Ewine F. van Dishoeck \orcidlink{0000-0001-7591-1907} \inst{6,7} \and
          Valentin Christiaens \orcidlink{0000-0002-0101-8814} \inst{8,12} \and
          Aditya M. Arabhavi \orcidlink{0000-0001-8407-4020} \inst{1} \and
          Thomas Henning \orcidlink{0000-0002-1493-300X} \inst{9} \and
          Manuel G\"udel \orcidlink{0000-0001-9818-0588} \inst{10,11} \and
          Peter Woitke \inst{2} \and
          Olivier Absil \orcidlink{0000-0002-4006-6237} \inst{12} \and
          David Barrado \orcidlink{0000-0002-5971-9242} \inst{13} \and
          Alessio Caratti o Garatti \orcidlink{0000-0001-8876-6614} \inst{14,15} \and
          Adrian M. Glauser\orcidlink{0000-0001-9250-1547} \inst{16} \and
          Fred Lahuis \inst{5} \and
          Silvia Scheithauer \orcidlink{0000-0003-4559-0721} \inst{9} \and
          Bart Vandenbussche \orcidlink{0000-0002-1368-3109} \inst{8} \and
          Danny Gasman  \orcidlink{0000-0002-1257-7742} \inst{8} \and
          Sierra L. Grant \orcidlink{0000-0002-4022-4899} \inst{7} \and
          Nicolas T. Kurtovic \orcidlink{0000-0002-2358-4796} \inst{7} \and
          Giulia Perotti \orcidlink{0000-0002-8545-6175} \inst{9} \and
          Beno\^{i}t Tabone \orcidlink{} \inst{17} \and
          Milou Temmink \orcidlink{0000-0002-7935-7445} \inst{6}
            }
\authorrunning{J.Kanwar et al.}
\institute{Kapteyn Astronomical Institute, University of Groningen, P.O. Box 800, 9700 AV Groningen, The Netherlands \email{kanwar@astro.rug.nl} \and
    Space Research Institute, Austrian Academy of Sciences, Schmiedlstr. 6, A-8042, Graz, Austria \and 
    TU Graz, Fakultät für Mathematik, Physik und Geodäsie, Petersgasse 16 8010 Graz, Austria \and
    Department of Astrophysics/IMAPP, Radboud University, PO Box 9010, 6500 GL Nijmegen, The Netherlands \and
    SRON Netherlands Institute for Space Research, Niels Bohrweg 4, NL-2333 CA Leiden, the Netherlands \and
    Leiden Observatory, Leiden University, 2300 RA Leiden, the Netherlands \and
    Max-Planck Institut f\"{u}r Extraterrestrische Physik (MPE), Giessenbachstr. 1, 85748, Garching, Germany \and 
    Institute of Astronomy, KU Leuven, Celestijnenlaan 200D, 3001 Leuven, Belgium \and
    Max-Planck-Institut f\"{u}r Astronomie (MPIA), K\"{o}nigstuhl 17, 69117 Heidelberg, Germany \and
    Dept. of Astrophysics, University of Vienna, T\"urkenschanzstr. 17, A-1180 Vienna, Austria \and
    ETH Z\"urich, Institute for Particle Physics and Astrophysics, Wolfgang-Pauli-Str. 27, 8093 Z\"urich, Switzerland \and
    STAR Institute, Universit\'e de Li\`ege, All\'ee du Six Ao\^ut 19c, 4000 Li\`ege, Belgium \and
    Centro de Astrobiolog\'ia (CAB), CSIC-INTA, ESAC Campus, Camino Bajo del Castillo s/n, 28692 Villanueva de la Ca\~nada, Madrid, Spain \and
    INAF – Osservatorio Astronomico di Capodimonte, Salita Moiariello 16, 80131 Napoli, Italy \and
    Dublin Institute for Advanced Studies, 31 Fitzwilliam Place, D02 XF86 Dublin, Ireland \and
    ETH Z\"urich, Institute for Particle Physics and Astrophysics, Wolfgang-Pauli-Str. 27, 8093 Z\"urich, Switzerland \and
    Universit\'e Paris-Saclay, CNRS, Institut d’Astrophysique Spatiale, 91405, Orsay, France
    }

 
  \abstract
  {With the advent of JWST, we acquire unprecedented insights into the physical and chemical structure of the inner regions of planet-forming disks where terrestrial planet formation occurs. The very low-mass stars (VLMS) are known to have a high occurrence rate of the terrestrial planets around them. Exploring the chemical composition of the gas in these inner regions of the disks can aid a better understanding of the connection between planet-forming disks and planets.}
  {The MIRI mid-Infrared Disk Survey (MINDS) project is a large JWST Guaranteed Time program to characterize the chemistry and physical state of planet-forming and debris disks. We use the JWST-MIRI/MRS spectrum to investigate the gas and dust composition of the planet-forming disk around the very low-mass star Sz28 (M5.5, 0.12\,M$_{\sun}$).}
  {We use the dust-fitting tool (DuCK) to determine the dust continuum and to get constraints on the dust composition and grain sizes. We use 0D slab models to identify and fit the molecular spectral features, yielding estimates on the temperature, column density and the emitting area. To test our understanding of the chemistry in the disks around VLMS, we employ the thermo-chemical disk model {P{\tiny RO}D{\tiny I}M{\tiny O}} and investigate the reservoirs of the detected hydrocarbons. We explore how the C/O ratio affects the inner disk chemistry.}
  {JWST reveals a plethora of hydrocarbons, including \ce{CH3}, \ce{CH4}, \ce{C2H2}, \ce{^{13}CCH2}, \ce{C2H6}, \ce{C3H4}, \ce{C4H2} and \ce{C6H6} suggesting a disk with a gaseous C/O\,>\,1. Additionally, we detect \ce{CO2}, \ce{^{13}CO2}, \ce{HCN}, and \ce{HC3N}. \ce{H2O} and OH are absent in the spectrum. We do not detect PAHs. Photospheric stellar absorption lines of \ce{H2O} and \ce{CO} are identified. Notably, our radiation thermo-chemical disk models are able to produce these detected hydrocarbons in the surface layers of the disk when the C/O\,>\,1. The presence of C, \ce{C+}, H and \ce{H2} are crucial for the formation of hydrocarbons in the surface layers and a C/O ratio larger than 1 ensures a surplus of C to drive this chemistry. Based on this, we predict a list of additional hydrocarbons that should also be detectable. Both amorphous and crystalline silicates (enstatite, forsterite) are present in the disk and we find grain sizes of 2 and 5\,$\mu$m.}
  {The disk around Sz28 is rich in hydrocarbons and has a high gaseous C/O ratio in the inner regions. In contrast, it is the first VLMS disk in the MINDS sample to show distinctive dust features together with a rich hydrocarbon chemistry. The presence of large grains indicates dust growth and evolution. Thermo-chemical disk models employing an extended hydrocarbon chemical network together with C/O\,>1 are able to explain the hydrocarbon species detected in the spectrum.}

   \maketitle
%

\section{Introduction}\label{Introduction}
The evolution of gas and dust around very low-mass stars (VLMS) \citep{Liebert1987} is only now starting to be extensively studied with new telescopes. It is important to study these objects since the occurrence of terrestrial planets around such stars is quite high. Transit observations find 2.5~$\pm$~0.2 planets per M dwarf with radii 1\,-\,4\,$R_{\oplus}$ and periods of $<200$\,days \citep{Dressing2015}, and radial velocity searches find $\sim$1.32 planets with 1\,$M_{\oplus}$\,<\,$M_{pl}$ sin $i$\,<\,10\,$M_{\oplus}$ and periods $<100$\,days \citep{Sabotta2021, Schlecker2022}.
Studying the inner regions of the planet-forming disks around M type stars sheds light on the chemistry and dust evolution in these very low-mass objects in a phase where planets are likely forming \citep{Henning2013}. With the advent of the James Webb Space Telescope (JWST) \citep{Rigby2023}, we probe the warm and dense inner regions (1\,au, $T$$\sim$200\,-\,1000\,K, $n$\,$\sim$10$^{10}$\,-\,10$^{18}$\,cm$^{-3}$) of disks around such very low-mass stars with a sensitivity and spectral resolution higher than the Spitzer Space Telescope thus giving insight in the gas and dust properties in this region. As part of the MINDS GTO collaboration \citep{Henning2024,Kamp2023}, the disks around the VLMS J1605321-1933159 \citep[J160532 here in after,][]{Tabone2023} and ISO-Chal 147 \citep{Arabhavi2023} have been reported to have high C/O ratios in their inner regions. This inference is based on the detection of a plethora of complex hydrocarbons. These authors do not detect any silicate feature in the spectra thus indicating dust growth and evolution as large grains tend to emit at longer wavelengths. Contrary to these sources there are disks around other VLMS such as Sz114 \citep{Xie2023} which are rich in water and only show the standard hydrocarbon \ce{C2H2} which has been previously detected with Spitzer in many T~Tauri and VLMS disks \citep{Salyk2008,Carr2011, Pontoppidan2010,Pascucci2009}. \cite{Xie2023} conclude that different initial conditions such as a large and massive disk and potential substructures may have kept water abundances high in the inner disk resulting in a water-rich spectrum.

\begin{figure*}[t]
   \centering
    \includegraphics[width=\linewidth]{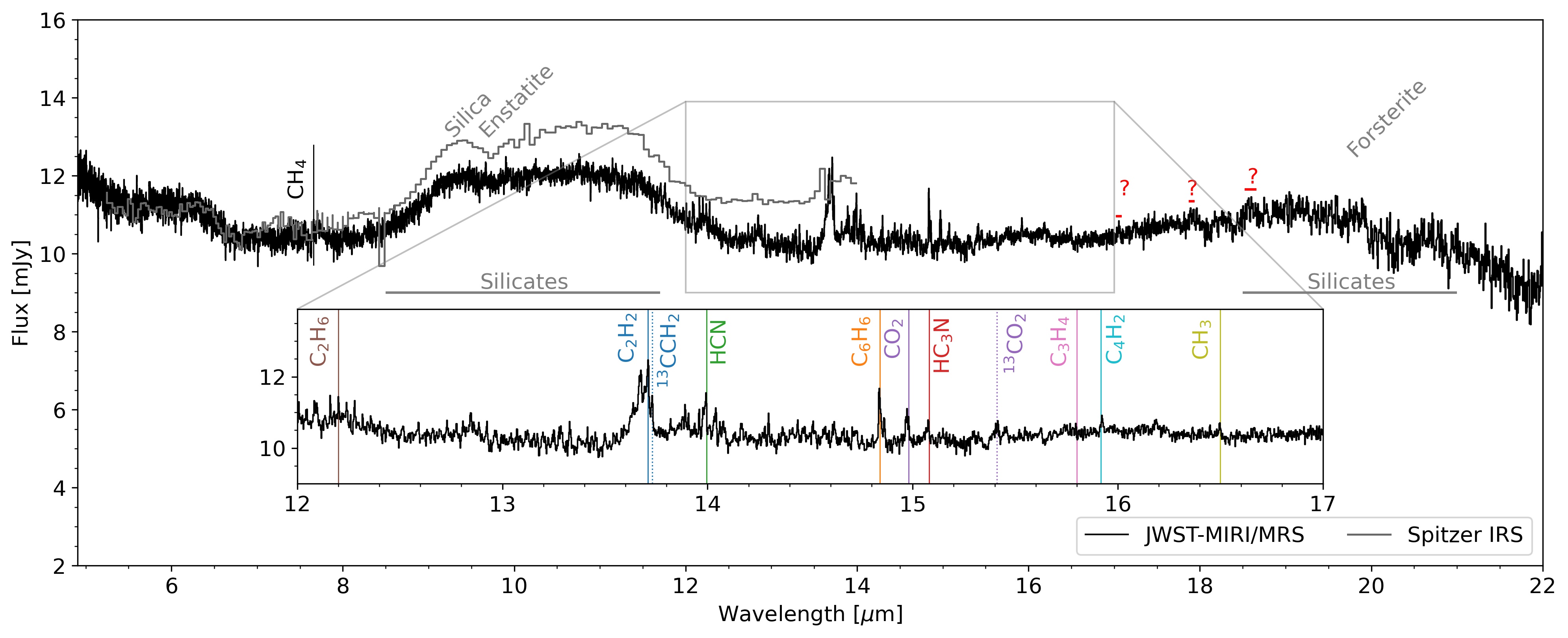}
   \caption{JWST-MIRI/MRS spectrum of Sz28 labelling the detected molecular species and the dust features. The spectrum from Spitzer is depicted in grey. The red question mark (?) symbol depicts the unidentified spectral features and the red horizontal lines show their wavelength range. The Q-branch of the molecules are shown in solid lines along with their identified isotopologues, if any, in dotted lines.}
   \label{main}%
\end{figure*}

We present here the JWST-Mid InfraRed Instrument (MIRI) \cite[]{miri_rieke2015PASP,Wright2015,Wright2023} Medium-Resolution Spectrometer (MRS) \cite[]{Wells2015,Argyriou2023} spectrum of the source Sz28 (T37 or 2MASS-J11085090-7625135) and investigate whether it is similar to the hydrocarbon rich sources J160532 and ISO-Chal 147 or to the water rich source Sz114. Sz28 is a very low-mass star \citep[M$_{\star}$\,=\,0.12\,M$_{\sun}$, L$_{\star}$\,=\,0.03\,L$_{\sun}$]{Manara2017}. We rescale L$_{\star}$ to 0.04\,L$_{\sun}$ based on the new distance from Gaia 3 \citep{Galli2021}. It is a member of the $\sim$3.6\,Myr old \citep{Ratzenb2023} Chamaeleon {\sc I} star-forming region at a distance of 192.2\,pc \citep{Galli2021}. Its spectral type is M5.5, it has an effective temperature of $\sim$\,3060\,K and an accretion rate of $\sim$1.8$\cdot$10$^{-11}$\,M$_{\sun}\,$yr$^{-1}$ \citep{Manara2017}. The ALMA non-detection sets an upper limit on the disk dust mass of 0.48\,M$_{\oplus}$ \citep{Pascucci2016} and on the gas mass of either 0.08\,M$_{\rm Jup}$ or 3.1\,M$_{\rm Jup}$ deduced using two different CO isotopologue methods \citep{Long2017}. Spitzer detected \ce{C2H2} line emission and revealed that the dust in this source is more processed compared to the interstellar medium dust exhibiting comet-like 9.4\,$\mu$m and 11.3\,$\mu$m crystalline silicates \citep{Pascucci2009}. 

Considerable effort focused on understanding the strong features of \ce{C2H2} in disks even before JWST. \cite{Agundez2008, Bast2013, Walsh2015} and \cite{Kanwar2023} investigated the chemistry of \ce{C2H2} in the warm high density gas of the inner disk and how its abundance structure changes with stellar properties and the use of different chemical networks and rate databases. \cite{Woods2007} and \cite{Kanwar2023} outlined the pathways to form benzene in planet-forming disks. We aim here to use thermo-chemical models to aid the interpretation of the spectrum and study the chemistry in the inner regions of the disks around VLMS. We explore the effect of a change in C/O elemental ratio on the hydrocarbon chemistry. 

The paper is structured in the following way. Section\,\ref{Observations and Data Reduction} describes the observations, the methods used to reduce the data and lists the molecular detections. We then describe the determination of the dust continuum and slab model retrievals used for the subsequent quantitative analysis in Sect.\,\ref{Analysis methods}. The non-detections, detection of dust features along with the findings from the slab models analysis such as gas temperature, column density and emitting area are reported in Sect.\,\ref{Results}. We compare our astro-chemical understanding to that of the species detected in the JWST-MIRI/MRS spectrum using thermo-chemical disk models with varying C/O ratio in Sect.\,\ref{Hydrocarbon chemistry in disks}. The comparison of Sz28 with other disks around VLMS in terms of its gas and dust content along with the implications of our results on planet formation are discussed in Sect.~\ref{Discussion}. Finally, we summarise our findings and conclusions in Sect.\,\ref{Conclusion}. 

\section{Observations and data reduction}\label{Observations and Data Reduction}

Sz28 was observed on August 16, 2022, at 02:06:10 for a total time of 1.02 hours in MRS mode with the JWST-MIRI instrument over the full wavelength range of 4.9$-$27.9~$\mu$m as part of Guaranteed Time Observation (GTO) program 1282 (PI: Th. Henning). It was observed with target acquisition in FASTR1 readout mode using a four-point dither pattern in the negative direction with an exposure time of 924 seconds in each band. The spectrum contains 4 channels, each with 3 bands.

We use version 1.11.1 \citep{Bushouse2023} of the JWST standard Science Calibration pipeline with the CRDS context jwst\textunderscore1094.pmap and VIP package \citep[version 1.4.2;][]{GomezGonzalez2017, Christiaens2023, Christiaens2024} for the reduction of the uncalibrated files. We used \texttt{Detector1} of the pipeline followed by stray-light correction when either the \texttt{sdither} or \texttt{ddither} background subtraction is used. We skip stray-light correction if we use the \texttt{annulus} background subtraction method; for more information see \cite{Christiaens2024}. The default parameters for \texttt{Spec2} were used. The VIP-based routines for bad pixel correction are used instead of the outlier detection in \texttt{Spec3}, as the former efficiently corrects for significant spikes that would otherwise affect the extracted spectrum of the source \citep[e.g.][]{Perotti2023}. As Sz28 had no dedicated background observation, we explored various background subtraction techniques. Initially, we used the \texttt{annulus} background subtraction where the median value in an annulus directly encircling the aperture used for photometry is subtracted. Subsequently, we tested the subtraction of a background proxy estimated from the rate files obtained with the four-point dither pattern, and smoothed with a median filter (\texttt{sdither}). Both of these methods resulted in MIRI flux levels comparable to that of Spitzer, but the spectrum was noisy. We finally used the direct pair-wise dither subtraction (\texttt{ddither}) to estimate the background as it led to a higher $S/N$ in the reduced spectrum. This method results in lower flux levels than the Spitzer observations. This minor flux discrepancy is attributed to the self-subtraction between minimally overlapping PSFs. Because of the higher $S/N$, this spectrum is then used for further analysis.

As there is no line-free region in the spectrum we use the JWST Exposure Time Calculator (ETC)\footnote{https://jwst.etc.stsci.edu/} to obtain an ideal estimate for the $S/N$ as described in \cite{Temmink2024}. This noise only takes into account the photon and read noise of the detector. The actual noise therefore, would be higher than we report. We use the ratio of median continuum flux level in each band and the estimated $S/N$ as the noise level in the spectrum. These values are reported in Table~\ref{ETC} for bands where we detect molecular emission. The estimated noise level for the Spitzer spectrum was $\sim$0.4\,mJy \citep{Pascucci2009} whereas the estimated noise level for the MIRI/MRS is $\sim$0.1\,mJy in channel 2. The flux differences are within the 10\% Spitzer flux calibration uncertainties and 5\% absolute spectro-photometric MRS uncertainty \citep{Argyriou2023}.

Figure\,\ref{main} presents the identified dust and molecular features in the JWST-MIRI/MRS spectrum of Sz28 along with the Spitzer low resolution spectrum for comparison. Note that many of the small wiggles seen in the spectrum at $<20\,\mu$m are not high frequency noise but are due to individual molecular lines.

\begin{table*}
\caption{Molecular detections in Sz28}
\begin{tabular}{lll} \hline
Molecule & Emission Band  & Wavelength ($\mu$m)\\ \hline
\ce{CH3}&  out-of-plane bending mode ($\nu_2$)& 16.48 \\
\ce{CH4}& degenerate deformation mode ($\nu_4$) & 7.65\\
\ce{C2H2}& asymmetric bending mode ($\nu_5$)  & 13.71 \\
$^{13}$CCH$_2$ & asymmetric bending mode ($\nu_5$) & 13.73 \\
\ce{C2H6} & \ce{CH3} rocking mode ($\nu_9$) & 12.17 \\
\ce{C3H4}\tablefootmark{a} & CH bending mode ($\nu_9$) & 15.80 \\
\ce{C4H2} & CH bending mode ($\nu_8$) & 15.92\\
\ce{C6H6} & out-of-plane bending mode ($\nu_4$) & 14.85\\ [5pt]
\multirow{2}{*}{\ce{CO2}} &
\multirow{2}{*}{\begin{tabular}[c]{@{}l@{}}fundamental bending mode ($\nu_2$)\\ excited bending mode ($\nu_1$\,$\nu_2$\,$\nu_3$: 100-010) \end{tabular}} &
\multirow{2}{*}{\begin{tabular}[c]{@{}l@{}}14.98\\ 13.88, 16.18\end{tabular}} \\
 &  &  \\ [5pt]
\multirow{2}{*}{$^{13}$\ce{CO2}} &
  \multirow{2}{*}{\begin{tabular}[c]{@{}l@{}}fundamental bending mode ($\nu_2$)\\ excited bending mode ($\nu_1$\,$\nu_2$\,$\nu_3$: 100-010)\end{tabular}} &
  \multirow{2}{*}{\begin{tabular}[c]{@{}l@{}}15.41\\ 16.20\end{tabular}} \\ & & \\ [5pt]
\multirow{2}{*}{HCN} &
  \multirow{2}{*}{\begin{tabular}[c]{@{}l@{}}fundamental bending mode ($\nu_2$:1-0)\\ excited bending mode ($\nu_2$:2-1)\end{tabular}} &
  \multirow{2}{*}{\begin{tabular}[c]{@{}l@{}}14.00\\ 14.30\end{tabular}} \\ & & \\ [5pt]
\ce{HC3N} & HCC bending mode ($\nu_5$) & 15.08 \\ \hline  
\end{tabular}
\tablefoot{\tablefoottext{a}{propyne, \ce{CH3CCH}}}
\label{emission}
\end{table*}

\begin{table}[]
\caption{The $S/N$ provided by ETC along with the median continuum flux in each band where molecules are detected.}
\resizebox{\linewidth}{!}{%
\begin{tabular}{llll}
\hline
Band [$\mu$m]   & Median Flux [mJy] & $S/N$ by ETC & $\sigma$ [mJy]\\ \hline
1B(5.66-6.63)   & 10.72            & 101     & 0.10 \\
1C(6.53-7.65)   & 9.87            & 114     & 0.08  \\
3A(11.55-13.47) & 10.0           & 116     & 0.08   \\
3B(13.34-15.57) & 9.40            & 104     & 0.09  \\
3C(15.41-17.98) & 10.11           & 100     & 0.10  \\ \hline
\end{tabular}}
\tablefoot{$\sigma$ is the estimated noise in each band.}
\label{ETC}
\end{table}

We consider a molecule as detected if we identify a Q-branch above the 3$\sigma$ level. For species that do not show a Q-branch, multiple P- and/or R-branch lines above 3$\sigma$ are required to conclude a detection. In this way we detect the molecular emission bands listed in Table\,\ref{emission}.
Several clear features indicated by red questions marks on the spectrum (Fig.~\ref{main}) remain so far unidentified. In the following, we will detail the quantitative methods used to analyse the spectrum further.

\begin{figure*}
   \centering
    \includegraphics[width=\linewidth]{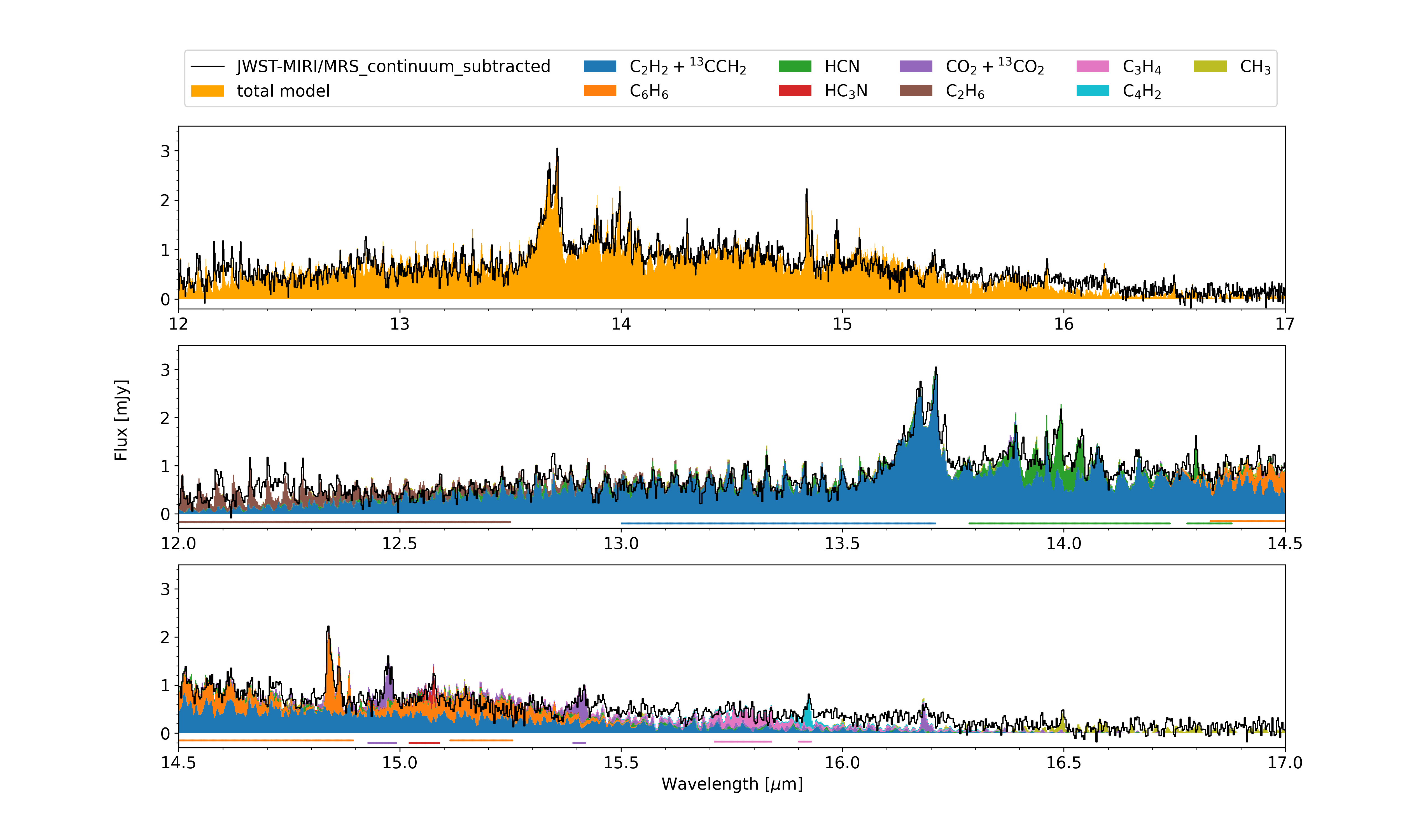}
   \caption{The top panel shows the continuum subtracted JWST-MIRI/MRS spectrum of Sz28 showing the total model (addition of slabs of all the detected molecules) in the wavelength region 12 to 17\,$\mu$m. The bottom two panels show the individual contribution of the slab models for the detected molecular species. The slab model fits are performed over the wavelength window shown at the bottom of the panel as straight line for a molecule in their corresponding color. The slab parameters of the spectrum of not well-constrained molecules can be found  
   in Table\,\ref{not_fit}.}
              \label{slabs}%
    \end{figure*}

\begin{figure}
   \centering
    \includegraphics[width=\linewidth]{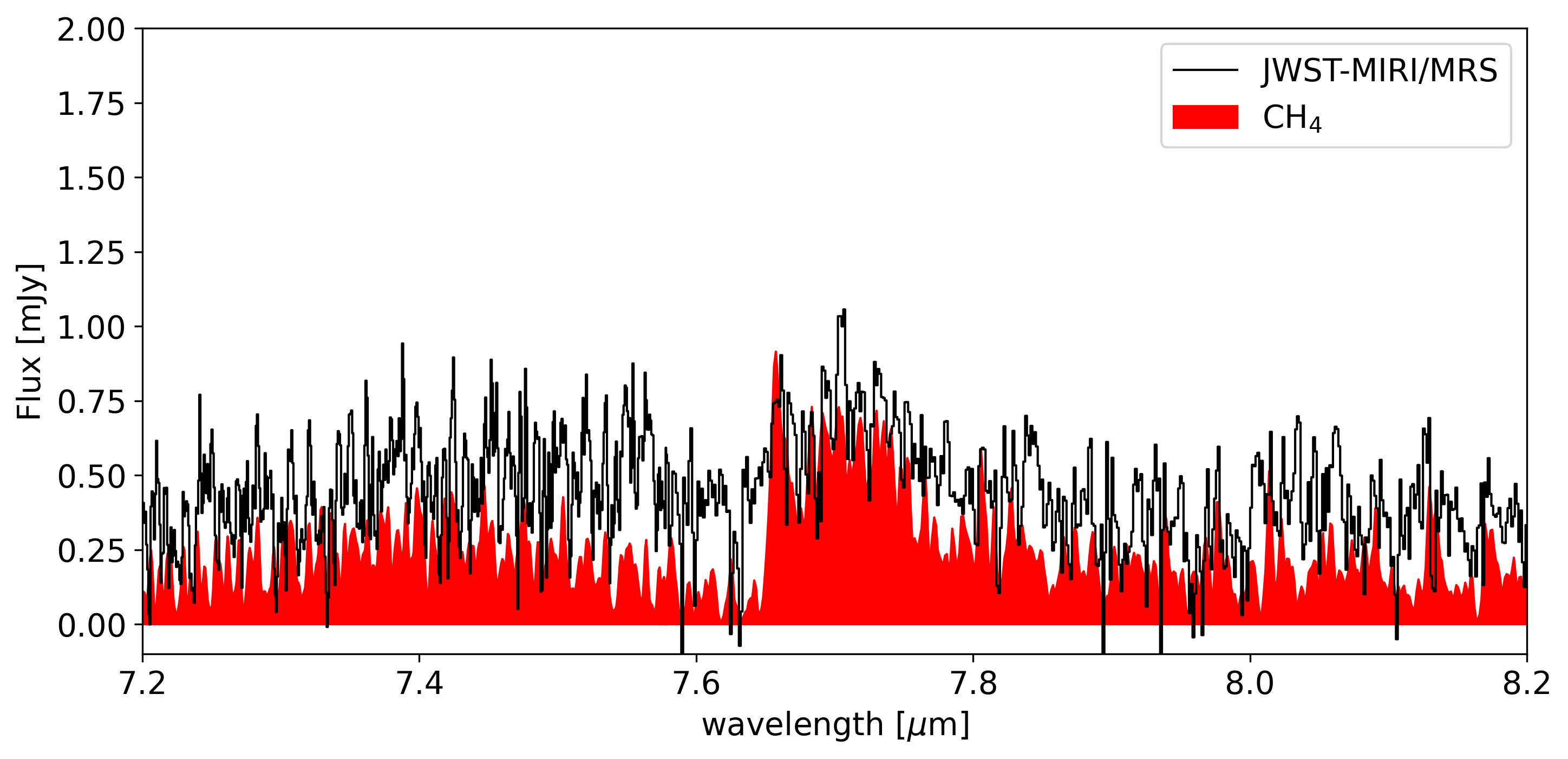}
   \caption{The slab model for \ce{CH4} is shown in red along with the zoomed-in JWST-MIRI/MRS spectrum of Sz28. The parameters from the 0D slab models for the represented slab can be found
   in Table\,\ref{not_fit}.}
              \label{CH4}%
    \end{figure}

\section{Analysis methods}\label{Analysis methods}

\subsection{Continuum subtraction}\label{resulting_continuum}

To identify the molecules in the spectrum, we need to determine a reliable dust continuum. In contrast with other VLMS, we observe the silicate dust feature at 10\,$\mu$m and an unusual, weak 16\,$\mu$m feature. We first assumed the lower envelope of the spectrum as the dust continuum. Based on the residuals of the initial fitting of the molecular features, the presence of a molecular pseudo-continuum was found similar to \cite{Arabhavi2023}. 

To subsequently better guide our continuum placement and explain the $\sim$16\,$\mu$m feature we employed the dust retrieval code (Dust Continuum Kit; DuCK) explained in detail in \cite{Kaeufer2023} and Jang et al.\ (in prep.), using a similar approach as \citet{Juh2010}. It is a 1D code that uses three components to explain the dust emission: a rim, optically thick midplane and optically thin surface layer and uses Bayesian analysis to fit the observations. The code uses a library of dust species. For crystalline dust, we use enstatite \citep{Jaeger1998}, and forsterite \citep{Servoin1973}. For amorphous dust, we use silica \citep{Spitzer1960}, silicates with pyroxene \citep{Dorschner1995} and olivine \citep{Henning1996} stoichiometry. We then fit the dust features based on these dust absorption coefficients using a range of grain sizes (0.1-5\,$\mu$m). The Gaussian random field method is used to calculate the Q curves for the dust species \citep[Jang et al.\ in prep.,][]{Min2007}.  

When using the entire spectrum, we find that the dominant dust species varies between the shorter and longer wavelengths. This is because the code gives more weight to shorter wavelengths resulting in a fit that failed to capture the overall shape of the dust continuum at longer wavelengths. Consequently, this makes it difficult to recover the peculiar feature at 16\,$\mu$m. To address this issue, we opted to divide the spectrum into two wavelength ranges, short (4.9-15\,$\mu$m) and long (15-22\,$\mu$m), and use the dust model to fit them separately. Fitting the spectrum in such a way, produces a forsterite feature around 16\,$\mu$m after taking into account the molecular emission contribution to the continuum. The peak of the observed feature is slightly blue-shifted compared to this dust model fit. However, this shift could be attributed to imperfect opacity curves. This feature could thus be explained by four different scenarios: different dust composition, imperfect opacity curves, a molecular pseudo-continuum or a combination thereof.

We deviate from the continuum found using the dust retrieval code based on the residuals that are due to molecular features. We also exclude the data longwards of 22\,$\mu$m as it is noisy and spurious due to lack of signal. As the dust continuum was overestimating the dust strength at various wavelengths, we selected certain points informed by the dust fit as shown in Fig.\,\ref{continuums}. We then use a cubic-spline fit function \citep[CubicSpline]{Dyer2001} for interpolating the continuum which is now only informed by the dust-fit models as shown in Fig.\,\ref{continuums} and is therefore, not the direct output of the dust retrieval code (see App.\,\ref{Figures depicting the continuum} for more details). 

\subsection{0D modelling}\label{0D modelling}

We use 0D slab models for the identification of the molecules and to estimate the gas temperature $T$, column density $N$ and the emitting radius $R_\mathrm{em}$ consistent with the observed molecular emissions. $T$ is varied in intervals of 25\,K from 25\,K to 1500\,K. $N$ in log-scale is varied from 10$^{14}$ to 10$^{24.5}$\,cm$^{-2}$ in intervals of 0.1$\overline{6}$\,dex. The emitting area in log-scale is varied between 0.01 and 100\,au. We use {P{\tiny RO}D{\tiny I}M{\tiny O}} \citep{Woitke2009, Woitke2016a} and the prodimopy (v2.1.4) Python package for generating slab models and performing $\chi^2$ fits, respectively, as explained in \citet{Arabhavi2023}. The spectroscopic data is taken from HITRAN \citep{Gordon2022}, \citealt[\ce{(C3H4,C6H6)}] {Delahaye2021} and \citealt[(\ce{CH3})]{Helmich1996}. For \ce{C3H4} and \ce{C6H6}, we use the partition functions, Einstein coefficients and degeneracies from \cite{Arabhavi2023}. The slab models are convolved to the MIRI/MRS resolving power ({R\,=\,}2500 for channel 3 and 3500 for channel 1) and resampled to the JWST-MIRI/MRS wavelength grid using the Python package SpectRes \citep{Carnall2017}. We use the noise estimate described in Sect.\,\ref{Observations and Data Reduction}. 
For \ce{C2H2} and \ce{CO2} whose emission spans bands 3A and 3B, and, 3B and 3C, respectively, we calculate the final noise ($\sigma$) as:
\begin{equation}
    \sigma_{\ce{C2H2}}=\frac{W_{\rm3A}\cdot \sigma_{\rm3A} + W_{\rm3B}\cdot \sigma_{\rm3B}}{W_{3A}+W_{3B}}
\end{equation}
\begin{equation}
    \sigma_{\ce{CO2}}=\frac{W_{\rm3B}\cdot \sigma_{\rm3B} + W_{\rm3C}\cdot \sigma_{\rm3C}}{W_{3B}+W_{3C}}
\end{equation}
where $W$ is the number of data points used to calculate $\chi^2$ from each band \citep{Temmink2024}.
We use the resulting continuum subtracted spectrum (see Sect.\,\ref{resulting_continuum}) to fit the slab models for one molecule or group of molecules at a time and subtract this best fit slab from the spectrum. The residual spectrum is then used for fitting the subsequent molecules, again one at a time as described below. We calculate the $\chi^2$ between the continuum subtracted observed spectrum and the convolved and re-sampled slab modeled spectrum \citep{Kamp2023,Grant2023}. The mask used in the $\chi^2$ determination for each molecule is defined over the wavelength region which has the least contamination from other molecules (see Fig.\,\ref{slabs}). 

We do not fit bands 1A (4.9\,$\mu$m\,-\,5.74\,$\mu$m) and 1B\,(5.66\,$\mu$m-6.63\,$\mu$m) as they are dominated by the stellar spectrum (see Fig.~\ref{star+disk}). The low resolution stellar model spectrum obtained from PHOENIX \citep{Brott2005} using the stellar parameters (Sect.\,\ref{Introduction}) is not good enough to correct for this; more work is required to obtain a good fitting high resolution stellar spectrum. We also do not fit any molecular emission features between 8.9 and 11.8\,$\mu$m as they are hard to disentangle from the dust emission.

Here we explain our fitting methodology (especially the order and grouping of molecules) in more detail as the spectrum shows blending of molecular emission at various wavelengths. We find channel 3B in the wavelength range 13.0\,-\,14.5\,$\mu$m to be dominated by the $\nu_5$ mode of \ce{C2H2}. The slab models for \ce{C2H2} include its isotopologue $\rm^{13}$CCH$_2$ in the ratio 1:35 \citep{Woods2009} and are thus treated as a \textquote{single} species. The spectroscopic data for $^{13}$CCH$_2$ are available only for the lowest $\nu_5$ (1-0) band. We use optically thick ($N$>10$^{19}\,\mathrm{cm}^{-2}$) and thin ($N$<10$^{19}\,\mathrm{cm}^{-2}$) slabs of \ce{C2H2} to fit the line emission in this band and treat it as fitting two species simultaneously 
\citep[thus the free parameters are $N_1$ x $T_1$ x $R_{\mathrm{em}1}$ x $N_2$ x $T_2$ x $R_{\mathrm{em}2}$ where 1 and 2 correspond to the optically thick and thin components, see][]{Arabhavi2023}. The $\chi^2$ is evaluated in the wavelength window 13.00\,-\,13.71\,$\mu$m. We observe the Q-branch peaks of $^{13}$CCH$_2$ at 13.69 and 13.73\,$\mu$m but purposely define the fitting window excluding the latter wavelength as it leads to over-prediction of the free parameters due to the incomplete spectroscopic data for $^{13}$CCH$_2$. We then subtract this best fit slab model (thick and thin components of \ce{C2H2}) from the continuum subtracted spectrum. Next, we perform the fit for \ce{C6H6} in the wavelength region 14.3\,-\,15.27\,$\mu$m, because this is now the dominant feature in the residual spectrum. Again, the resulting best fit slab for \ce{C6H6} is subtracted from the spectrum. In a next step, the new residual spectrum is then used to fit the remaining molecules one by one in the following order \ce{HC3N}\,-\,\ce{CO2}\,-\,\ce{HCN}\,-\,\ce{C2H6} (only for confirming the detection)\,-\,(\ce{C4H2}\,+\,\ce{C3H4}; fitted together). 
The wavelength windows over which the $\chi^{\rm2}$ is evaluated to determine best fits are performed are shown in Fig.\,\ref{slabs}. We do not fit \ce{CH3} as we have incomplete spectroscopic data comprising of only a single band for this molecule; we use this only for confirming the detection. In band 1C, 6.5-8.2\,$\mu$m, we only have a clear detection of \ce{CH4} (see Fig.\,\ref{CH4}). We do not fit the species since the presence of absorption lines suggests likely a stellar contribution which we cannot easily correct for. Figure\,\ref{CH4} shows a slab model visually matched to the data of \ce{CH4} to depict its detection.


\section{Results}\label{Results}

The subsequent paragraphs describe more quantitatively the dust and molecular content that we find in the JWST-MIRI/MRS spectrum. 

\subsection{Dust composition and sizes}
\label{Sect:DuCK}

The shape of the continuum indicates the presence of silicate dust. The shape and strength of the silicate feature is determined by the grain size and composition \citep{Henning2010}. Based on the dust fitting that decomposes the spectrum into emission from dust species, we identify the 9.2$\,\mu$m feature of \ce{SiO2}, the 11.3 and 19.3\,$\mu$m 
features of forsterite, and the 9.4, 9.9, 18.2\,$\mu$m features of enstatite. Thus, the dust composition comprises amorphous and crystalline silicates which indicates that the dust in the disk has undergone thermal processing. The mass fraction of crystalline silicates is $\sim$\,20\% and the rest are amorphous silicates. The ratio of the two crystalline dust species enstatite and forsterite change between the wavelength regions. Forsterite is more abundant than enstatite in the longer wavelength region (18.8\% and 3\%, respectively), whereas in the shorter wavelength region, the abundances of forsterite and enstatite are similar (9.4\% and 7.5\%, respectively). For amorphous silicates, the dust model indicates quite large grains, typically 5\,$\mu$m, and for forsterite, we find both large grains $\sim$5\,$\mu$m and slightly smaller grain sizes, typically 2\,$\mu$m. The dust models predict systematically smaller grain sizes $\sim$\,2\,-3\,$\mu$m for enstatite. 

\subsection{Gas molecular content}

Figure\,\ref{slabs} shows the resulting continuum subtracted spectrum and the 0D slab model fits for all the detected species. We provide the values of $T$, $N$ and $R_\mathrm{em}$ consistent with these molecular emission in Table\,\ref{Table} along with 1\,$\sigma$ uncertainities. We note that there is a degeneracy between the gas temperature $T$ and the column density $N$ for the same species, which is evident from the $\chi^2$ plots (see Fig.\,\ref{chi2maps}). The $S/N$ of the spectrum is low and subsequently, many fits are not well constrained. 

The slab model fitting indicates that the optically thick \ce{C2H2} has high column densities giving rise to a molecular pseudo-continuum. This optically thick emission originates from a smaller emitting area compared to the optically thin emission similar to the findings of \cite{Tabone2023} and \cite{Arabhavi2023}. \ce{C6H6} has its temperature constrained between 110\,K and 550\,K with column densities ranging between 10$^{17.5}$\,-\,10$^{19.5}\,\mathrm{cm}^2$.

The temperatures and column densities of the molecules \ce{CH3}, \ce{CH4}, \ce{HCN}, \ce{HC3N}, \ce{CO2}, \ce{C2H6}, are poorly constrained and therefore, we refrain from providing 'best fit parameters' in Table\,\ref{Table}. The parameters describing the slabs used to identify these species in Fig.\,\ref{slabs} are listed in Table\,\ref{not_fit}. \ce{CH3} is not fitted as we have incomplete spectroscopic data. HCN is detected, but its emission is blended with \ce{C2H2} and therefore it is hard to constrain its parameters. The emission of \ce{HC3N} and \ce{CO2} overlaps with the P-branch of \ce{C6H6} making it difficult to provide best fit values. We find that \ce{CO2} is cold compared to many of the other molecules with temperatures between 100\,K and 250\,K and column densities greater than 10$^{18.5}\,\mathrm{cm^{-2}}$. This low temperature is driven  by the narrow shape and the peak positions of the Q-branches of the main and hot $^{12}$\ce{CO2} bands and the $^{13}$\ce{CO2} band (see Fig.~\ref{CO2_blowup}).
\ce{HC3N} has a column density lower than 10$^{21}\,\mathrm{cm^{-2}}$ and the temperature varies between 50\,K and 690\,K as evident from the $\chi^2$ map in Fig\,\ref{chi2maps}. \ce{C2H6} is also poorly constrained in our analysis. Its spectral signature coincides with the region where the spectral leak correction was applied (at 12.2\,$\mu$m \citealt{Wells2015}). Unfortunately, the 1.11.1 version of the standard pipeline did not yet correct for this leak. 

There are still many unidentified spectral features longward of 16.5\,$\mu$m, some of which are indicated with a question mark in Fig\,\ref{main}. In addition, we were not able to identify the 12.844\,$\mu$m broad feature. 

We do not detect any PAH emission features, OH, \ce{NH3}, atomic or molecular hydrogen emission or metal fine-structure lines given the current quality of the spectrum. We do not detect \ce{H2O} or CO emission from the disk. This is because the short wavelength region up to 6.5\,$\mu$m is dominated by the stellar spectrum as seen in Fig.\,\ref{star+disk}. The overall shape of the PHOENIX stellar spectrum matches well with the observed Sz28 MIRI/MRS spectrum, although the stellar model has a lower spectral resolution than MIRI/MRS. We clearly detect \ce{H2O} and \ce{CO} stellar absorption lines similar to the case of ISO-Chal 147 \cite[][Fig.\,S6]{Arabhavi2023}. To search for the potential presence of \ce{H2O} or CO emission from the disk, we thus first require high resolution photospheric stellar models to disentangle the stellar and disk contribution at the short wavelengths. This is outside the scope of this paper.

The richness in hydrocarbon features and non-detection of oxygen-bearing molecules (except \ce{CO2}) point towards an inner disk with an elemental ratio C/O$\,>\,1$. The absence of the [Ne\,\textsc{II}] line is in agreement with the upper limit of stellar X-rays log($L_{\mathrm{X}}$/erg\,s$^{-1}$)\,$\sim$28.9 \citep{Gudel2010}. The non-detection of \ce{H2O}, OH and CO does not necessarily imply their absence, but rather puts an upper limit to their column densities and/or emitting areas given the estimated noise level (0.1\,mJy).

\begin{table}[]
\caption{Best fit slab model parameters for the molecules detected in Sz28.}
\resizebox{\linewidth}{!}{%
\begin{tabular}{lllll}
\hline
Molecules  & $T$ (K)  & log$_{10}$\,$N$ (cm$^{-2}$) & $R_\mathrm{em}$ (au) & $\mathcal{N}$\\ \hline 
\\
\multicolumn{1}{l}{\multirow{2}{*}{C$_2$H$_2$}} & \multirow{2}{*}{425}$^{\mathrm{+175}}_{\mathrm{-225}}$ & 20.67$_{\mathrm{-2.17}}$     & 0.015$^{\mathrm{+0.015}}_{\mathrm{-0.005}}$ \\[5pt]
\multicolumn{1}{l}{}                            &                      & 18.33$^{\mathrm{+1.67}}_{\mathrm{-1.03}}$     & 0.03$^{\mathrm{+0.07}}_{\mathrm{-0.02}}$  \\ [5pt]
C$_6$H$_6$                                      & 225$^{\mathrm{+325}}_{\mathrm{-115}}$                  & 18.17$^{\mathrm{+1.33}}_{\mathrm{-0.67}}$    & 0.07$^{\mathrm{+0.53}}_{\mathrm{-0.06}}$  & 5$\times$10$^{42}$ \\ [5pt]
C$_3$H$_4$                                      & 250$^{\mathrm{+200}}_{\mathrm{-50}}$                  & 15.83$^{\mathrm{+1.67}}_{\mathrm{-0.83}}$      & 0.38$^{\mathrm{+0.62}}_{\mathrm{-0.33}}$  & 1.2$\times$10$^{41}$ \\[5pt]
C$_4$H$_2$                                      & 250$^{\mathrm{+200}}_{\mathrm{-50}}$                  & 16.33$^{\mathrm{+1.67}}_{\mathrm{-0.83}}$     & 0.067$^{\mathrm{+0.93}}_{\mathrm{-0.017}}$ & 6.7$\times$10$^{40}$\\ [5pt]
\hline
HCN                                             & detected                  &      &        \\
CO$_2$                                          & detected                  &        &      \\
HC$_3$N                                         & detected                  &     &         \\
C$_2$H$_6$                                      & detected                  &      &        \\
CH$_3$                                          &  detected            &           &     \\ 
CH$_4$ & detected      &        & \\ \hline
\end{tabular}}
\tablefoot{For optically thin emission, $\mathcal{N}$ represents the number of molecules obtained as a product of $N \times \pi R_{\mathrm{em}}^{\rm{2}}$. The values for the optically thick and thin component of \ce{C2H2} are provided. The fitting procedure did not yield a well-constrained solution for the molecules labelled as \textquote{detected}.}
\label{Table}
\end{table}

As explained above, in  many cases $T$, $N$ and $R_\mathrm{em}$ for molecules are not well-constrained. However, all these molecules except \ce{CO2} share some common parameter space indicating that they might be sharing a single emitting reservoir. \ce{CO2} is the only molecule with temperatures as low as 100-250\,K based on the $\chi^2$ maps, indicating that \ce{CO2} could have a different emitting region. We also followed the approach of \cite{Tabone2023} to fit species using the emitting area of the optically thick or thin component of \ce{C2H2} to minimize uncertainties and better constrain parameters $T$ and $N$. However, using either of the emitting areas, we were not able to explain the flux levels observed in the spectrum for other molecules such as \ce{C6H6}. Simultaneously fitting the dust and molecules in a wavelength region and allowing for radial density/temperature gradients may constrain the parameter space better, but is outside the scope of this paper. It requires advanced fitting strategies such as CLIcK \citep{Liu2019} or DuCKLinG \citep{Kaeufer2023}.

\section{Hydrocarbon chemistry in disks around VLMS}\label{Hydrocarbon chemistry in disks}

The above analysis has shown that the gas in the inner disk is very rich in hydrocarbons, leaving open the question whether our chemical networks can adequately describe the formation of all these hydrocarbons in such warm, high density disk environments. The 0D slab models used in the Sect.\ref{0D modelling} do not consider temperature gradients, dust opacity, the gas heating/cooling balance, chemistry, and are therefore not suited to discuss and understand the chemical pathways at play.
To investigate the above question, we use radiation thermo-chemical disk models. 
Our primary objective here is not to fit the observed spectrum, but to test whether the hydrocarbon molecules we detect are consistent with our understanding of the chemistry in these environments (high density inner disks, high C/O).

\subsection{Thermo-chemical modelling}\label{Thermo-chemical modelling}

We use the radiation thermo-chemical code {P{\tiny RO}D{\tiny I}M{\tiny O}} to simulate a typical disk around a very low-mass star. This is a 2D code that can model the physical and chemical structure of the disk self-consistently \citep{Woitke2016a}. It calculates the dust temperature structure from dust radiative transfer and the gas temperature by balancing the gas heating and cooling processes. \cite{Greenwood2017} presented a brown dwarf {P{\tiny RO}D{\tiny I}M{\tiny O}} disk model with stellar properties similar to those of Sz28. We adopt the same disk model and change the stellar parameters to those of Sz28. A detailed list of the disk model parameters can be found in Table~\ref{Parameter}. To have a gradual build up in the surface density in the inner regions we use a soft inner rim as explained in \cite{Woitke2023}. We define the surface density $\Sigma$(r) as
\begin{equation}
    \Sigma(r) \propto \mathrm{exp}\,\Bigg[-\Bigl(\frac{R_{\rm soft}}{r}\Bigl)^{s}\Biggr]\, r^{-\epsilon} \, \mathrm{exp}\,\Bigg[-\Bigl(\frac{r}{R_{\rm tap}}\Bigl)^{2-\epsilon}\Biggr]
\end{equation}
where $R_\mathrm{tap}$ is the tapering-off radius,
\begin{equation}
    R_{\rm soft}\,=\,\mathrm{raduc}\,\times R_{\rm in}
\end{equation}
and
\begin{equation}
s\,=\,\frac{\mathrm{ln}\Big[\mathrm{ln(1/reduc)}\Bigr]}{\mathrm{ln(raduc)}}
\end{equation}
The power law exponent $\epsilon$ is 1. The raduc parameter defines the radius up to which the density is gradually building up in units of inner disk radius $R_\mathrm{in}$ and is chosen to be 1.15. The reduc parameter determines the reduction of the column density around the inner rim and is chosen to be 10$^{-7}$. We use the extended hydrocarbon chemical network added to the large DIANA chemical network \citep{Kamp2017, Kanwar2023} and steady state chemistry. The largest hydrocarbon species that this chemical network can form is \ce{C8H5+}. The reactions are primarily taken from UMIST2012 \citep{McElroy2013}. In addition, the network takes into account three body and thermal decomposition reactions. The elemental abundances are the low-ISM values taken from \cite{Kamp2017}. The gas and dust temperature and the UV radiation field calculated from detailed 2D radiative transfer (in units of the Draine field) of both the models are shown in Fig\,\ref{properties}.

We use two models to compare our chemical understanding to the molecular detections in our JWST mid-IR spectra of disks around VLMS. One model has the canonical C/O elemental ratio of 0.45 \citep{Anders1989}. The second model has a lower oxygen abundance, resulting in a C/O of 2. In both models, we keep the gas temperature structure fixed to that of the canonical model. This approach allows us to isolate the effects of the change in elemental abundance on the species abundances from the effect of gas temperature as the chemistry and gas temperature are naturally intertwined.

\subsection{Hydrocarbon chemistry}\label{Hydrocarbon chemistry}

The hydrocarbon chemistry has been recently revisited in the context of disks around T~Tauri stars \citep{Kanwar2023}.
They find that the dissociation of CO unlocks C (UV radiation) or \ce{C+} (X-rays or cosmic rays). This carbon can then abstract H from \ce{H2} or combine with another C to form hydrocarbons in the disk surfaces. However, in cases where oxygen is depleted, there is unblocked C already present to form hydrocarbons. The main neutral-neutral and ion-molecule pathways of formation and destruction for \ce{C2H2} and \ce{C6H6} in a disk around T~Tauri star are outlined in \cite{Kanwar2023}. These pathways agree with \cite{Walsh2015} for \ce{C2H2} and \cite{McEwan1999} for \ce{C6H6}. While \cite{Woods2007} find the ion-neutral pathway of \ce{C3H4} reacting with its own ion a major reaction leading to \ce{C6H6} via \ce{C6H7+}, \cite{Kanwar2023} find the reaction of \ce{C6H5+} with \ce{H2} via \ce{C6H7+} to dominate the \ce{C6H6} formation.

\begin{figure*}
   \centering
    \includegraphics[width=\linewidth]{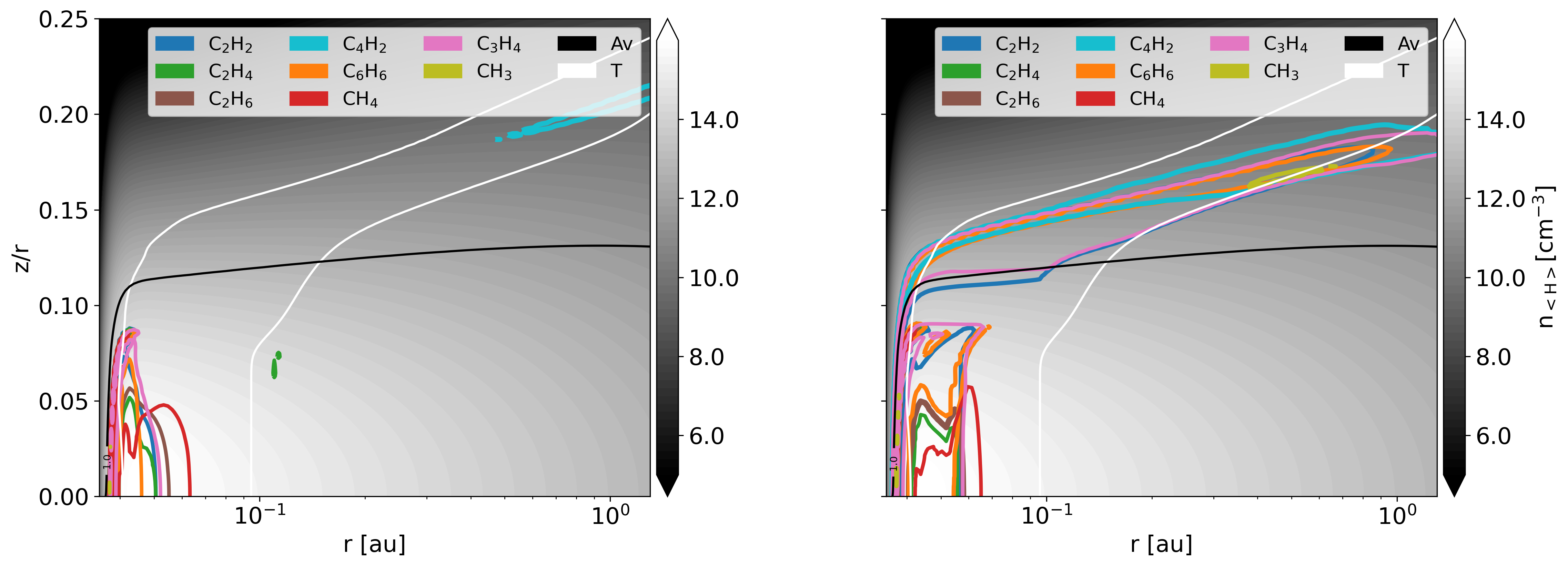}
   \caption{Abundance contours corresponding to 10\% of the maximum abundance of the hydrocarbons detected in disks around very low-mass stars in the canonical and enhanced model with C/O of 0.45 and 2.0, respectively. The white contours correspond to the gas temperatures of 140\,K and 475\,K. The A$_v$\,=1 mag contour is shown in black.}
              \label{together}%
\end{figure*}

We use here the above described model of a disk around a VLMS to investigate whether the gas phase formation and destruction pathways remain the same as for the T~Tauri disk and whether they depend on the C/O ratio. 
Figure\,\ref{together} shows where the observationally detected hydrocarbons reside in the disk for the canonical and enhanced C/O elemental ratio. The abundance contours outline a value of 10\% of their respective maximum abundance. Table~\ref{max_col_density} lists the maximum abundance, maximum total column density and total column density at 0.1\,au of these species in both models. Note that the maximum values are similar in both models, but the spatial region over which a species exhibits these corresponding values radially and vertically expands in the enhanced C/O model. Note that the molecular emission in the mid-IR spectrum is in fact limited by the dust continuum, and so does not reflect the full molecular column density (see Fig.\,\ref{properties}). This model is not tailored towards Sz28 specifically as we have too little constraints on the physical disk parameters. Thus, deep integration ALMA observations are required to constrain these parameters and improve the chemical understanding extracted from such thermo-chemical disk models. 

In the model with the canonical C/O elemental ratio, the detected hydrocarbon molecules are concentrated in a radially narrow, but vertically extended area in the inner 0.1\,au, close to the disk midplane. Increasing the C/O ratio to 2, the chemistry shifts to a carbon-dominated chemistry. This results in increased abundances by $\sim$4 to $\sim$10 orders of magnitude in the surface layers and in the inner disk, respectively (see Figs.~\ref{Abundance_plots} and \ref{Abundance_plots2}). 

We next analyse whether there are new chemical pathways in both models that become active leading to such complex hydrocarbons; we use \ce{C2H2} and \ce{C6H6} as proxies to be able to compare with our earlier work on T\,Tauri disks \citep{Kanwar2023}. We use here the same approach and analyse the grid point $r$\,=\,0.248\,au and $z$\,=\,0.042\,au lying in the line emitting region of \ce{C2H2} to study the chemical pathways of \ce{C2H2} and \ce{C6H6}. This point is characterized by the following condition: $T_\mathrm{gas}$\,=\,270\,K and $T_\mathrm{dust}$\,=\,230\,K, $n_\mathrm{<H>}$\,=\,6.4$\cdot \rm10^{10}cm^{-3}$, $A_\mathrm{V}^{ver}$\,=\,0.018, vertical visual extinction, $A_\mathrm{V}^{rad}$\,= 1.1, radial visual extinction. We find that the chemical pathways leading to \ce{C2H2} in the canonical model (disk around a VLMS, C/O=0.45) are similar to the findings of \cite{Kanwar2023} (disk around a T~Tauri star, C/O=0.45).
We find that the reaction 
\begin{equation} \label{R1}
    \ce{C2H} + \ce{H2} \rightarrow \ce{C2H2} + \ce{H}
\end{equation}
contributes $\sim$\,91\% to the formation of \ce{C2H2} in the canonical C/O model and $\sim$\,96\% in the model with enhanced C/O. The photo-dissociation of \ce{C2H2} to \ce{C2H} contributes $\sim$\,71\% in the model with canonical C/O and $\sim$\,37\% to the destruction of \ce{C2H2} in the carbon enhanced model. The chemistry in the canonical C/O model is driven by oxygen, but when this element is depleted, other reactions having carbon as one of the reactants become more important for the formation of \ce{C2H2} (see App.\,\ref{Chemistry in model with canonical and enhanced C/O ratio} for more details). A high C/O ratio leads to high abundances of hydrocarbons in the disk surface layers and those destroy \ce{H2} through H abstraction. This  leads to the H/\ce{H2} transition being pushed slightly deeper in the disk compared to the canonical C/O model (see Fig\,\ref{H_H2} and App.~\ref{Chemistry in model with canonical and enhanced C/O ratio} for more details).

\ce{C6H6} is formed in both VLMS disk models primarily via
\begin{equation}
    \ce{C6H7+} + \ce{e-} \rightarrow \ce{C6H6} + \ce{H}
\end{equation}
It is the only formation reaction available in UMIST2012. \ce{C6H7+} is formed via \ce{C6H5+} in both models similar to the findings of \cite{Kanwar2023}. \ce{C6H5+} is formed via radiative association between \ce{C2H2} and \ce{C4H3+} in both models. The abundance of \ce{C2H2} increases in the model with enhanced C/O which subsequently, results in an increased abundance of \ce{C6H6}. The reactions contributing to the destruction of \ce{C6H6} shuffle in importance because of the increased \ce{C2H2} abundance. Appendix\,\ref{Chemistry in model with canonical and enhanced C/O ratio} provides an in-depth analysis. 

In general, we find that the dominant formation and destruction pathways do not change for these two species with increasing C/O. The reaction rates increase in the model with an enhanced C/O ratio and we find a shift in the relative importance of the pathways. We did not find any \textquote{new} pathways that become active only when increasing the C/O ratio. However, we note that the details of the pathways depend on the location in the disk due to the gas temperature and densities changing throughout the disk.

This generic model for a disk around a VLMS also predicts the following hydrocarbons to be co-spatial with the detected hydrocarbons: \ce{C2}, \ce{C2H}, \ce{C3}, \ce{CH2CCH}, both cyclic and linear isotopomers of \ce{C3H2}, \ce{C5H2}, \ce{CH3C4H}, \ce{C6}, \ce{C6H2}. Unfortunately, none of these hydrocarbons have ro-vibrational mid-IR spectra in the HITRAN \citep{Gordon2022} or GEISA \citep{Delahaye2021} databases. However, the pure rotational lines of the two species \ce{C2H} and \ce{c-C3H2} have been detected in the mm and sub-mm wavelength regions with ALMA in disks around T~Tauri stars albeit at much lower column densities \citep{Qi2013, Bergin2016}. These molecules may also have observable ro-vibrational lines. Theoretical calculations along with laboratory spectroscopic measurements are needed to search for these molecules in the mid-IR spectra. 

\section{Discussion}
\label{Discussion}

\begin{table*}[]
\caption{Properties of the VLMS disks observed with JWST-MIRI/MRS.}
\label{VLMS}
\begin{tabular}{lllll} \hline
Properties                   & Sz28             & Sz114\tablefootmark{a}             & J1605321\tablefootmark{b}         & ISO-Chal 147\tablefootmark{c} \\ \hline
$M_{\star}$\,[$M_{\sun}$]    & 0.12             & 0.16              & 0.16             & 0.11\\
$L_{\star}$\,[$L_{\sun}$]    & 0.04             & 0.196             & 0.04             & 0.01 \\
$T_{\rm eff}$\,[$K$]           & 3060             & 3022\tablefootmark{d}              & 3126             & 3060 \\
Spectral type                & M5.5             & M5                & M4.75            & M5.75  \\
Star-forming region          & Chamaeleon {\sc I}     & Lupus {\sc III}         & Upper Sco        & Chamaeleon {\sc I} \\
Evolutionary Age\,[$Myr$]\tablefootmark{e}    & 3.5              & 6               & 3-19              & 3.5 \\
$\dot{M}_{\rm{acc}}$\,[$M_{\sun}\,$yr$^{-1}$] & 1.8$\times$10$^{-11}$ & 7.9$\times$10$^{-10}$ & 10$^{-10}$\,-\,10$^{-11}$ & 7.0$\times$10$^{-12}$\\
$L_\mathrm{\rm acc}\,[L_\sun$] & 8.9$\times$10$^{-5}$ & 1.99$\times$10$^{-3}$ & 2.57$\times$10$^{-4}$\tablefootmark{f} & 4.5$\times$10$^{-5}$ \\
$L_\mathrm{X}$\,[erg s$^{-1}$]  & 7.9$\times$10$^{28}$\tablefootmark{g} & 8.9$\times$10$^{29}$\tablefootmark{h}  &1.1$\times$10$^{29}$  & 1.1$\times$10$^{29}$ \\
$\bigl[\rm{Ne}\,\textsc{II}\bigr]$ & $\times$ & $\checkmark$ & $\times$ & $\times$ \\ \hline                        
\end{tabular}
\tablefoot{\tablefoottext{a}{properties from \cite{Xie2023}}\tablefoottext{b}{properties from \cite{Tabone2023}}\tablefoottext{c}{properties from \cite{Arabhavi2023}}\tablefoottext{d}{ \cite{Manara2023}}\tablefoottext{e}{\cite{Ratzenb2023}} \tablefoottext{f}{\cite{Pascucci2013}}\tablefoottext{g}{\cite{Gudel2007}}\tablefoottext{h}{ \cite{Gondoin2006}}.}
\end{table*}

\subsection{Dust evolution}\label{dust}

We find that Sz28 has clear distinctive dust features. The almost flat 10\,$\mu$m silicate feature indicates significant grain growth consistent with the fast dust evolution (1-3\,Myrs) in VLMS as shown in \cite{Apai2005}. The large grain sizes obtained from our dust models (see Sect.\,\ref{Results}) also align with the grain growth scenario. The grains seem systematically larger than for T~Tauri disks, where winds could explain the depletion of sub-micron sized grains \citep{Olofsson2009, Oliveira2011}. 

\cite{Apai2005} studied the dust processing in disks around brown dwarfs by studying the link between the strength and shape of silicate features. They analysed the flux ratio at 11.3 and 9.8\,$\mu$m against the peak flux over continuum ratio. We get a value of 1 for the former and 1.4 for the latter. This places Sz28 in \citet[][Fig.\,2]{Apai2005} in a region where the correlation between weak dust feature and crystalline silicates contribution is non-linear meaning the crystalline silicate features become dominant. The crystalline mass fraction of Sz28 ($\sim$\,20\%) is consistent with the findings of \cite{Apai2005} for VLMS (9\,-\,48\%). The presence of the crystalline dust features indicates thermal processing of the dust \citep{Pascucci2009}. 
Following \citet{Kessler2007}, the silicate emission region scales with the stellar luminosity as L$_\star^{0.56}$, suggesting that we probe the very inner regions of these VLMS disks, $\sim$0.05\,au.
Crystallization of dust can occur by thermal annealing at $\sim$700\,K or by gas-phase condensation of silicates at high temperatures \citep{Fabian2000}. Such temperatures can exist in the inner disk. These crystalline silicates can subsequently be redistributed in the disk by radial drift, diffusion or vertical turbulence \citep{Jang2024}.

Sz28 is not detected with ALMA in the continuum contrary to Sz114, and therefore is probably compact in dust. This could be due to an efficient radial drift, which is known to be stronger in disks around VLMS \citep[see ][]{pinilla2023}. There has been observational evidence of such strong radial drifts in bright disks around VLMS \citep{Kurtovic2021}. The high C/O ratio in the inner regions of the disk around Sz28 could indicate that significant radial drift has taken place \citep{Mah2023}. \cite{Mah2023} found that short viscous timescales and close-in icelines in VLMS result in a high gaseous C/O after 2\,Myr. \cite{Mah2024} take into account the substructures also. They introduce a gap at 10\,au beyond the \ce{CO2} iceline in their model at different times ranging from 0 to 1\,Myr. They find that for low-viscosity (<10$^{-4}$), the inner disk can remain abundant in water for long periods of time (atleast 5\,Myrs) irrespective of the depth of the gap and when it is formed (see \citealt[Figure\,D.2, top two panels]{Mah2024}). This may explain water-rich sources such as Sz114. However, with a viscosity of 3$\times 10^{-4}$, the inner disk can remain water-rich for up to 2\,Myrs and then begins to transition to a high C/O ratio.

Sz28 shows large grain sizes and some fraction of crystalline dust (based on the analysis of the Si feature in Sect.~\ref{Sect:DuCK}). The visual inspection of the silicate feature shows that Sz114 may have low crystalline dust fraction compared to Sz28 and the SED slope beyond 10\,$\mu$m is rising in the former where as it is flat in the latter. This would imply that Sz114 could have more \textquote{primitive} dust, indicating an earlier phase of evolution compared to Sz28. If the grains evolve as stated above, the presence of silicate features in Sz28 could indicate an intermediate phase of evolution while ISO-Chal 147 and J160532 are representative of later phases of evolution because they lack any evidence for a silicate feature (grain sizes grown beyond $\sim\,5~\mu$m or dust entirely depleted).

\subsection{Gas content of VLMS disks}

Sz28 shows a larger variety of hydrocarbons than J160532 whose spectrum is dominated by \ce{C2H2}. All the molecules detected in ISO-Chal 147 are also detected in Sz28 except for \ce{C2H4}. Similar to ISO-Chal 147, Sz28 also shows the 'magic'\footnote{'magic' refers to the more energetically stable species as reported in \cite{Sergey2022}} molecule composition. Similar to the other two sources, \ce{C2H2} also forms a molecular pseudo-continuum in Sz28 along with the presence of the isotopologue \ce{^{13}CCH2}. Due to the incompleteness of the spectroscopic data for $^{13}$CCH$_2$ and the noise level of the MIRI/MRS spectrum of this source, we cannot reliably determine the isotopic ratio. The molecules in common in the three disks are \ce{C2H2}, \ce{C4H2}, \ce{C6H6}, \ce{CO2} and \ce{^{13}CO2}. None of the three disks show PAHs, \ce{H2O} or OH emission. This suggests a high C/O ratio in the inner regions of all three disks. This is contrary to Sz114 that shows clear water and CO emission, likely indicating a low C/O ratio.

In Sz28, we find high column densities of hydrocarbons even in the presence of silicate features. Both ISO-Chal 147 and J160532 do not show any dust features and have high column densities of hydrocarbons. Thus, the absence of dust features does not necessarily imply detecting more hydrocarbons because we probe deeper into the disk.

All three sources show \ce{CO2} and strong \ce{C2H2} emission lines together; this may be hard to reconcile with an overall high C/O ratio,  where O would be predominantly bound in CO. \cite{Woitke2018} shows that the emission of \ce{CO2} is mainly coming from the tenuous surface layer rather than the inner reservoir. This may mean that \ce{CO2} does not necessarily need a high abundance to emit in the mid-IR (see Fig.\,\ref{Abundance_plots4}). 

\citet{Tabone2023} invoke the presence of a gap that holds back O-rich icy grains to  explain the high C/O in the inner disk. If we follow this idea, then it means that the outer disk could still have a normal C/O ratio and give rise to the \ce{CO2} emission. Such a scenario could explain why we systematically detect both species in VLMS. However, in that case, the \ce{CO2} emission should be systematically cooler than the hydrocarbons, but the degeneracies on $T$ and $N$ in the J160532 spectrum limits our conclusions \citep{Tabone2023}. 
Clearly, more work is needed to understand the simultaneous presence of strong \ce{C2H2} and \ce{CO2} emission in these disk spectra.

All the three disks differ from the water-rich disk around the VLMS Sz114. \cite{Xie2023} argue that the young disk has a supply of water due to the efficient radial drift of icy pebbles beyond the snowline. This disk may represent conditions which are a precursor to Sz28. 
Despite sharing similar spectral types, these objects show differences in dust content and molecular complexity and form the starting point of a deeper investigation using a larger sample of these objects.

The steady state gas-phase chemistry with an extended chemical network is able to form the species detected in the surface layers of disks around VLMS with a C/O\,>\,1 (depleted oxygen). This increase in the C/O ratio in the disk can be caused by various reasons, one being the close-in ice lines and the short viscous timescales ($\sim$2\,Myr) of disks around VLMS \citep{Mah2023}, the destruction of carbon grains \citep{Tabone2023, Arabhavi2023} or transport processes carrying O-rich ices on grains inside the snowline leading to O-rich gas that then gets accreted onto the star leaving behind a high C/O ratio gas \citep{Arabhavi2023}.
Based on the models, the timescales for the chemistry are $\sim$1\,yr, which means that the chemistry can adjust to the dynamic disk evolution which operates on viscous and grain drift timescales. Hence, we could conclude from this that the chemistry adjusts very quickly to any change in C/O ratio in the gas caused by any of the processes mentioned and our chemical network is adequate in describing the gas phase processes that occur therein.

\subsection{Radiation field environment}

Table\,\ref{VLMS} lists a number of key properties of the four VLMS that have been analysed so far. The estimated upper limit on the X-ray luminosity of Sz28 along with the absence of the [Ne\,\textsc{II}] and [Ne\,\textsc{III}] lines in the spectrum indicate a weak X-ray radiation field and/or the absence of high-velocity jets. This is similar to the other hydrocarbon-rich disks, J160532 and ISO-Chal 147. However, it is in contrast to Sz114 which has a relatively high X-ray luminosity (see Table\,\ref{VLMS}) and where the [Ne\,\textsc{II}] line indicates the presence of jets \citep{Xie2023}. It also has a higher $\dot{M}_{acc}$ and this is consistent with detection of \ce{H2} lines, similar to J160532. The low $L_\mathrm{acc}$ in Sz28 and ISO-Chal 147, is consistent with the non-detections of H\textsc{I} and/or \ce{H2} emission in both sources. J160532 on the other hand has stronger $L_{\rm{acc}}$ and indeed shows strong H\textsc{I} and \ce{H2} lines \citep{Francheschi2024}. 

The low UV field and the presence of dust in Sz28 also indicate that the carbon enrichment via photo-destruction \citep{Arabhavi2023} is less probable. The non-detection of PAHs can be either attributed to a lack of excitation due to the weak UV field or their absence in the disk.

\subsection{Evolutionary age}

Taking the stellar luminosity as a proxy for the age, \cite{Xie2023} argue that Sz114 is younger than J1605\textbf{32} and this could explain the low C/O ratio in the former following the disk evolution models from \cite{Mah2023}. Following this line of argument, the luminosity of ISO-Chal 147 being the lowest of the four indicates that it could be the oldest source. This would also be consistent with the conclusion drawn from the 10~$\mu$m dust emission feature (see Sect.\,\ref{dust}).

However, when assuming the ages of the star-forming regions (Table\,\ref{VLMS}) as a proxy for the individual sources, we would conclude that Sz28 and ISO-Chal 147 are of the same age, followed by Sz114 and J160532 would be the oldest one. Hence, this leaves us with an inconsistent picture concerning the evolutionary sequence of these four sources and a larger sample of VLMS is required to understand what drives the differences seen in the spectra (Arabhavi et al., in prep).

\subsection{Implications for planet formation}

Terrestrial planets form in dense, warm, inner regions of disks around these young stars. With JWST we probe these regions around Sz28 and detect plenty of hydrocarbons implying a high gaseous C/O ratio. The composition of the planets is influenced by their formation environment \citep{Paul2022,Oberg2011}, in this case, a high gaseous C/O ratio environment.

If the high C/O ratio in the gas reflects a similar C/O in the refractory material of the dust, it can then lead to future planets that have a layer of graphite on top of the silicate mantle \citep{Hakim2019}. However, if the dust is complimentary to the gas and thus has a low C/O ratio, it can instead lead to the formation of terrestrial planets that are carbon-poor like Earth.

The short wavelengths of the spectrum (below 6.5$\mu$m) are dominated by the stellar photosphere which shows the presence of O-bearing species like \ce{CO} and \ce{H2O} that hints at a low stellar C/O ratio. However, planet formation models that use observed stellar abundances show that the stellar photospheric C/O ratio is likely not correlated with the terrestrial planet's C/O \citep{Thiabaud2015} 

Protoplanets formed around VLMS can grow by pebble accretion \citep{Liu2020}. \cite{David2003,David2007} and \cite{Luhman2012} suggested that the apparent lifetime of disks are long ($\sim$10\,Myr). If the protoplanet grows sufficiently in size prior to the disk dispersal, it can hold its primordial atmosphere, which would be dominated by hydrogen and helium. This atmosphere can reflect the enrichment in volatiles (C, O, N, P) beyond H and He in the inner disk \citep{Krijt2022}. In hot evolutionary stages, the retention of the primordial atmosphere affects the evolution of magma. The solidification of magma is dependent on the primary volatiles in the atmosphere \citep{Lichtenberg2021}. The elemental composition thus inherited from the disk may play a role in the evolution of the planet, albeit minor and more indirectly. Thus, understanding the reservoirs of these detected molecules can provide additional insight into the composition of the potential terrestrial planets forming around these VLMS. A more detailed study using bulk composition and potentially atmosphere compositions of observed exoplanets around M-dwarfs could be very interesting in the future. 

\section{Conclusions}
\label{Conclusion}

We present and investigate the JWST-MIRI/MRS spectrum of the disk around the very low-mass star Sz28. We provide estimates for the temperature, column densities and the emitting area for a few of the observed species (see Table\,\ref{Table}). We then use thermo-chemical disk models to compare our astro-chemical understanding to these observed species. We present the reservoirs of the detected hydrocarbons and study the effect of a change of the C/O ratio on their abundances. Here we describe our final conclusions:

\begin{enumerate}
    \addtolength\itemsep{1mm}
    \item We identify the 9.2$\,\mu$m feature of \ce{SiO2}, the 11.3 and 19.3\,$\mu$m features of forsterite, and the the 9.4, 9.9, 18.2\,$\mu$m features of enstatite. We conclude that the grains in the disk surface have typical sizes of 2 to 5\,$\mu$m. Given the current data reduction technique and fitting the dust features up to 22\,$\mu$m, we find a significant mass fraction (>10\%) of crystalline silicates.
    
    \item We have firm detections of \ce{CH3}, \ce{CH4}, \ce{C2H2}, \ce{^{13}CCH2}, \ce{C2H6}, \ce{CO2}, \ce{^{13}CO2}, \ce{HCN}, \ce{HC3N}, \ce{C6H6}, \ce{C3H4} and \ce{C4H2}. We do not detect H\textsc{I}, \ce{H2}, \ce{OH}, \ce{H2O} or PAHs at longer wavelengths. At short wavelengths, the spectrum shows stellar photospheric \ce{CO} and \ce{H2O} absorption. We do not detect CO emission from the disk. Based on the type of molecules detected, we conclude that the disk has a carbon-dominated chemistry with a C/O ratio larger than 1. Based on the uncertainties in the parameters estimated with slab modelling, these molecules might share a common emitting reservoir except for \ce{CO2}.

    \item We find that the abundances of the hydrocarbons increase by 4 to 10 orders of magnitude when depleting oxygen (C/O$\,>\,2$) in thermo-chemical disk models. They appear in the surface layers and extend up to $\sim\,1$\,au. When analysing the steady-state gas-phase chemistry in the surface layers, we find a shift in the relative significance of the chemical pathways forming or destroying \ce{C2H2} and \ce{C6H6} when depleting oxygen. We do not identify any new pathways making substantial contributions to the chemistry. We find that the H/\ce{H2} transition layer is pushed slightly deeper in the disk due to hydrogenation of hydrocarbons which destroys \ce{H2}. The formation/destruction pathways of \ce{C2H2} and \ce{C6H6} are similar in disks around VLMS and T~Tauri stars.

    \item There are still several unidentified features in the MIRI/MRS spectrum beyond 17.0\,$\mu$m. From the thermo-chemical disk modelling, we find the following species co-spatial with the observationally detected molecules: \ce{C2}, \ce{C2H}, \ce{C3}, \ce{CH2CCH}, both cyclic and linear isotopomers of \ce{C3H2}, \ce{C5H2}, \ce{CH3C4H}, \ce{C6}, \ce{C6H2}. We need mid-IR spectra of these species to be able to check for their presence in the JWST spectra.
\end{enumerate}
The next step is to identify which processes lead to such an enhanced C/O ratio. A homogeneous sample of VLMS disks such as those studied with MINDS can serve as an excellent starting point for identifying observed trends and understanding processes that may be responsible for such high C/O ratios. A sample will also allow to investigate differences in chemical properties of the disk with respect to the stellar properties of VLMS. 

\begin{acknowledgements}
The authors acknowledge useful discussion with Wim van Westrenen. This work is based on observations made with the NASA/ESA/CSA James Webb Space Telescope. The data were obtained from the Mikulski Archive for Space Telescopes at the Space Telescope Science Institute, which is operated by the Association of Universities for Research in Astronomy, Inc., under NASA contract NAS 5-03127 for JWST. These observations are associated with program \#1282. The following National and International Funding Agencies funded and supported the MIRI development: NASA; ESA; Belgian Science Policy Office (BELSPO); Centre Nationale d’Etudes Spatiales (CNES); Danish National Space Centre; Deutsches Zentrum fur Luft- und Raumfahrt (DLR); Enterprise Ireland; Ministerio De Econom\'ia y Competividad; Netherlands Research School for Astronomy (NOVA); Netherlands Organisation for Scientific Research (NWO); Science and Technology Facilities Council; Swiss Space Office; Swedish National Space Agency; and UK Space Agency.
I.K. and J.K. acknowledge funding from H2020-MSCA-ITN-2019, grant no. 860470 (CHAMELEON). J.K. acknowledges support from SPFE.
I.K., A.M.A., and E.v.D. acknowledge support from grant TOP-1 614.001.751 from the Dutch Research Council (NWO).
E.v.D. acknowledges support from the ERC grant 101019751 MOLDISK and the Danish National Research Foundation through the Center of Excellence ``InterCat'' (DNRF150). 
T.H. acknowledge support from the European Research Council under the Horizon 2020 Framework Program via the ERC Advanced Grant Origins 83 24 28. 
D.B. has been funded by Spanish MCIN/AEI/10.13039/501100011033 grants PID2019-107061GB-C61 and No. MDM-2017-0737. 

A.C.G. acknowledges from PRIN-MUR 2022 20228JPA3A “The path to star and planet formation in the JWST era (PATH)” and by INAF-GoG 2022 “NIR-dark Accretion Outbursts in Massive Young stellar objects (NAOMY)” and Large Grant INAF 2022 “YSOs Outflows, Disks and Accretion: towards a global framework for the evolution of planet forming systems (YODA)”.

G.P. gratefully acknowledges support from the Max Planck Society.

Benoit:
B.T. is a Laureate of the Paris Region fellowship program, which is supported by the Ile-de-France Region and has received funding under the Horizon 2020 innovation framework program and Marie Sklodowska-Curie grant agreement No. 945298.

O.A. and V.C. acknowledge funding from the Belgian F.R.S.-FNRS.

D.G. and B.V. thank the Belgian Federal Science Policy Office (BELSPO) for the provision of financial support in the framework of the PRODEX Programme of the European Space Agency (ESA).

M.T. acknowledges support from the ERC grant 101019751 MOLDISK. We thank the anonymous referee for their constructive comments.

\end{acknowledgements}
\bibliographystyle{aa}
\bibliography{Papers}

\appendix
\twocolumn

\section{Continuum determination}
\label{Figures depicting the continuum}

\begin{figure}[!h]
   \centering    \includegraphics[width=\linewidth]{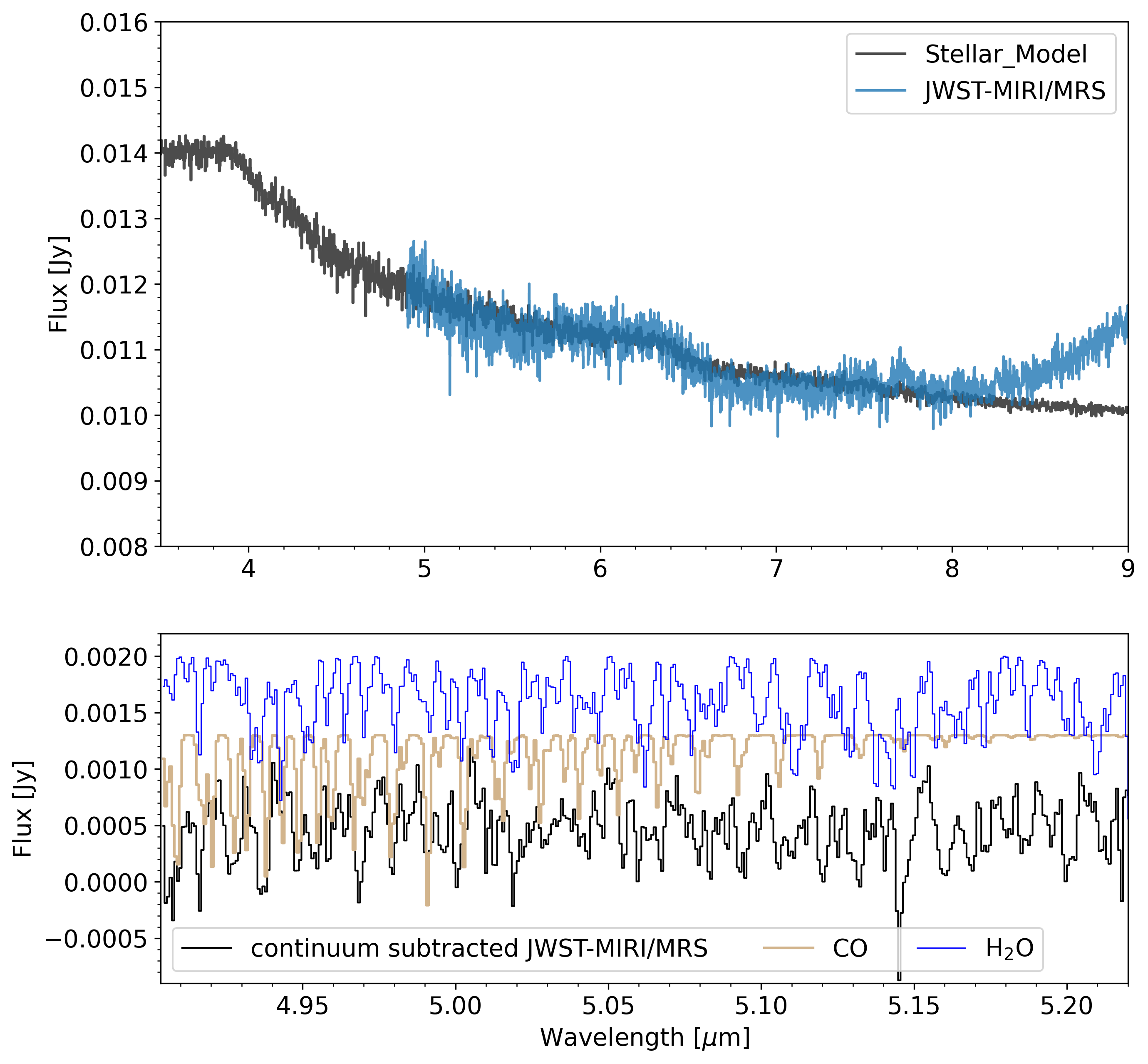}
   \caption{The top panel displays the JWST-MIRI/MRS spectrum of Sz28 and the modelled stellar spectrum which has an offset of 0.009\,Jy. The bottom panel depicts the CO and \ce{H2O} absorption slabs and the continuum subtracted JWST-MIRI/MRS spectrum highlighting their respective presence.}
              \label{star+disk}%
    \end{figure}

Figure\,\ref{star+disk} shows the zoomed in JWST-MIRI/MRS spectrum along with the modelled stellar photospheric spectrum of Sz28. The overall shape of the modelled stellar spectrum agrees well with the JWST-MIRI/MRS spectrum in channel 1A (4.90-5.73\,$\mu$m) and 1B (5.66-6.63\,$\mu$m). This indicates that the spectrum is dominated by the host star at the short wavelengths. The modelled stellar spectrum is at a lower resolution than MIRI/MRS. This hinders the removal of stellar contribution from the JWST-MIRI/MRS spectrum. The bottom panel of Fig.\,\ref{star+disk} shows the continuum-subtracted JWST-MIRI/MRS with the absorption slab models of CO and \ce{H2O} highlighting the presence of these molecules in the spectrum.

\begin{figure*}[!h]
   \centering
   \includegraphics[width=\linewidth]{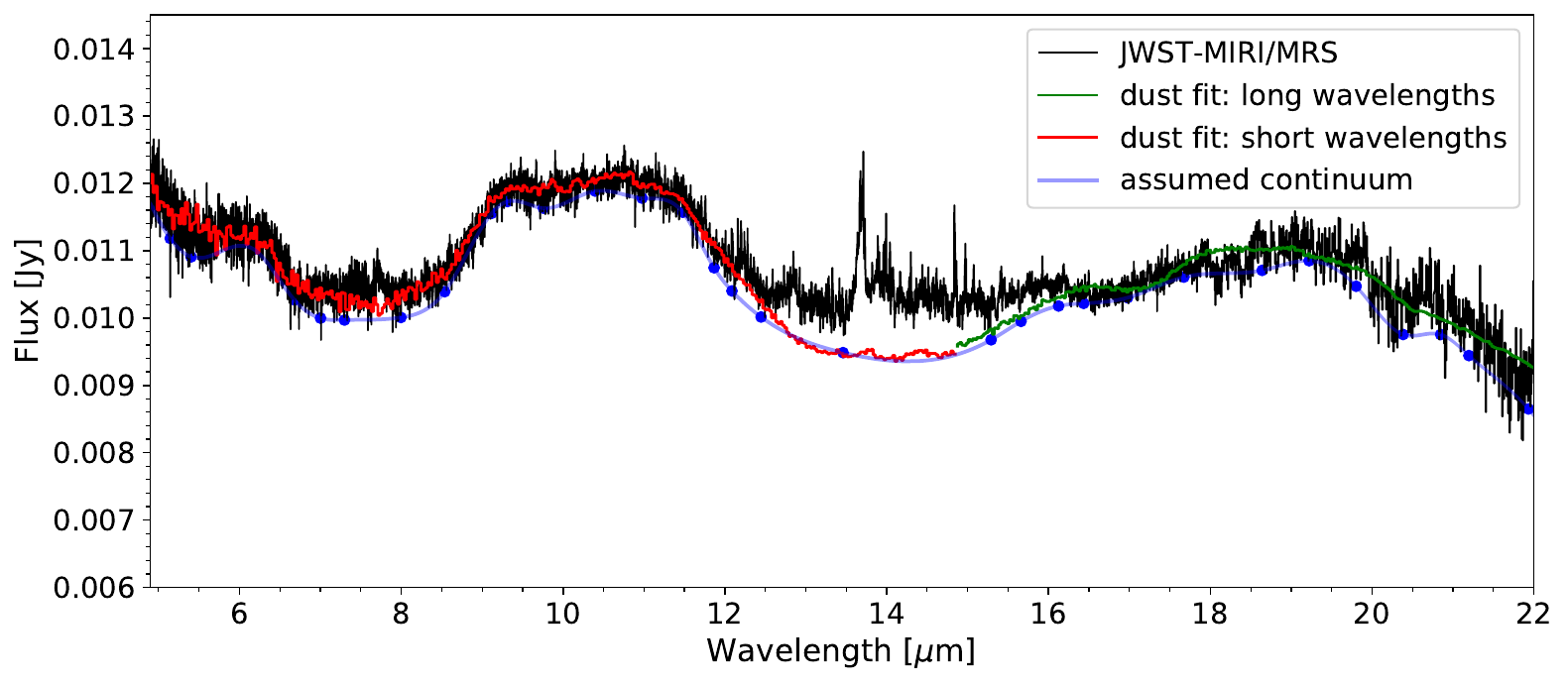}
   \caption{The interpolated dust continuum assumed (blue) together with the JWST-MIRI/MRS spectrum of Sz28 (black). The blue dots represent the locations used for interpolation. The continuum provided by the dust fits for the short and long wavelength part of the spectrum are shown in red and green, respectively.}
              \label{continuums}%
    \end{figure*}

Figure\,\ref{continuums} shows the continuum obtained from the dust fit models and the final continuum assumed for the slab model analysis. We describe here in some more detail how we arrived at this continuum using an iterative procedure. We first used a manually placed continuum to get an estimate of the molecular contribution. Subsequently, this molecular contribution between 12-17\,$\mu$m is removed from the spectrum based on this manually placed continuum. The resulting spectrum without the modeled molecular contribution between 12-17\,$\mu$m is then used for the dust-fitting tool DuCK \citep{Kaeufer2023} to find a refined dust continuum. This refined dust continuum is then adjusted according to the molecular contribution based on the residuals of the slab model fits. However, the resulting dust continuum was still not able to capture the overall shape. As stated in Sect.\,\ref{resulting_continuum}, the spectrum was then divided into a short and long-wavelength region. This provided an extra degree of freedom to adequately fit the overall shape. The dust compositions thus obtained show that the short wavelength region has similar mass fraction of enstatite and forsterite while the long wavelength region is forsterite dominated. 

\section{$\chi^2$ for molecules not quantitatively analysed}
\label{chi2 for molecules detected in T37}

For several molecules where we attempted 0D slab model fits, we find that the remaining degeneracies in the column density and temperature are too large to warrant a quantitative analysis.
We present in this  appendix the $\chi^2$ maps of these species. They are labelled as \textquote{detected} in Table~\ref{Table}. We do not provide such maps for \ce{CH3} and \ce{CH4} because we do not fit these molecules due to limitations defined in Sect.\,\ref{0D modelling}. We provide the $\chi^2$ map for \ce{C6H6} to illustrate the large error bars on the molecule that is relatively well-constrained.

We use $\chi^2_{min}$\,+\,2.3, $\chi^2_{min}$\,+\,6.2 and  $\chi^2_{min}$\,+\,11.8 as 1$\sigma$, 2$\sigma$ and 3$\sigma$ confidence intervals \citet[Table\,1 and Equation\,6]{Avni1976}. The black contours represent the emitting radii of 0.01, 0.05, 0.1, 0.5, 1, 5\,au. Table\,\ref{not_fit} provides the parameters of the slabs shown in Fig.\,\ref{slabs} for the detected molecules that are not well-constrained and hence not analysed quantitatively. It also provides the total number of molecules $\mathcal{N}$ for species that are optically thin.

\begin{table}[!h]
\caption{The parameters of the 0D slab models used to show the \textquote{detected} molecules in the JWST-MIRI/MRS spectrum of Sz28 along with number of molecules $\mathcal{N}$.}
\begin{tabular}{lllll}
\hline
Molecule            & $T$\,(K) & log$_{10}$\,$N$\,(cm$^{-2}$) & $R_\mathrm{em}$\,(au) & $\mathcal{N}$\\ \hline
HCN                 & 825       & 14.67     & 0.465  & 1$\times$10$^{\rm41}$        \\
\ce{HC3N}           & 250       & 15.5     & 0.159   & 2.4$\times$10$^{\rm40}$     \\
CO$_2$+$^{13}\mathrm C$O$_2$ & 100       & 21.67     & 0.824  &          \\
\ce{C2H6}           & 475       & 19.3      & 0.025   &       \\
\ce{CH3}            & 175       & 21        & 0.09    &     \\ 
\ce{CH4}            & 450       & 20        & 0.05    &      \\ \hline
\end{tabular}
\tablefoot{These parameters do not report the best fits but are rather utilised to show confirmed detection of the molecular emissions in the Sz28 spectrum.}
\label{not_fit}
\end{table}

\begin{figure*}[!h]
\begin{tabular}{lllll}
\includegraphics[width=0.3\textwidth]{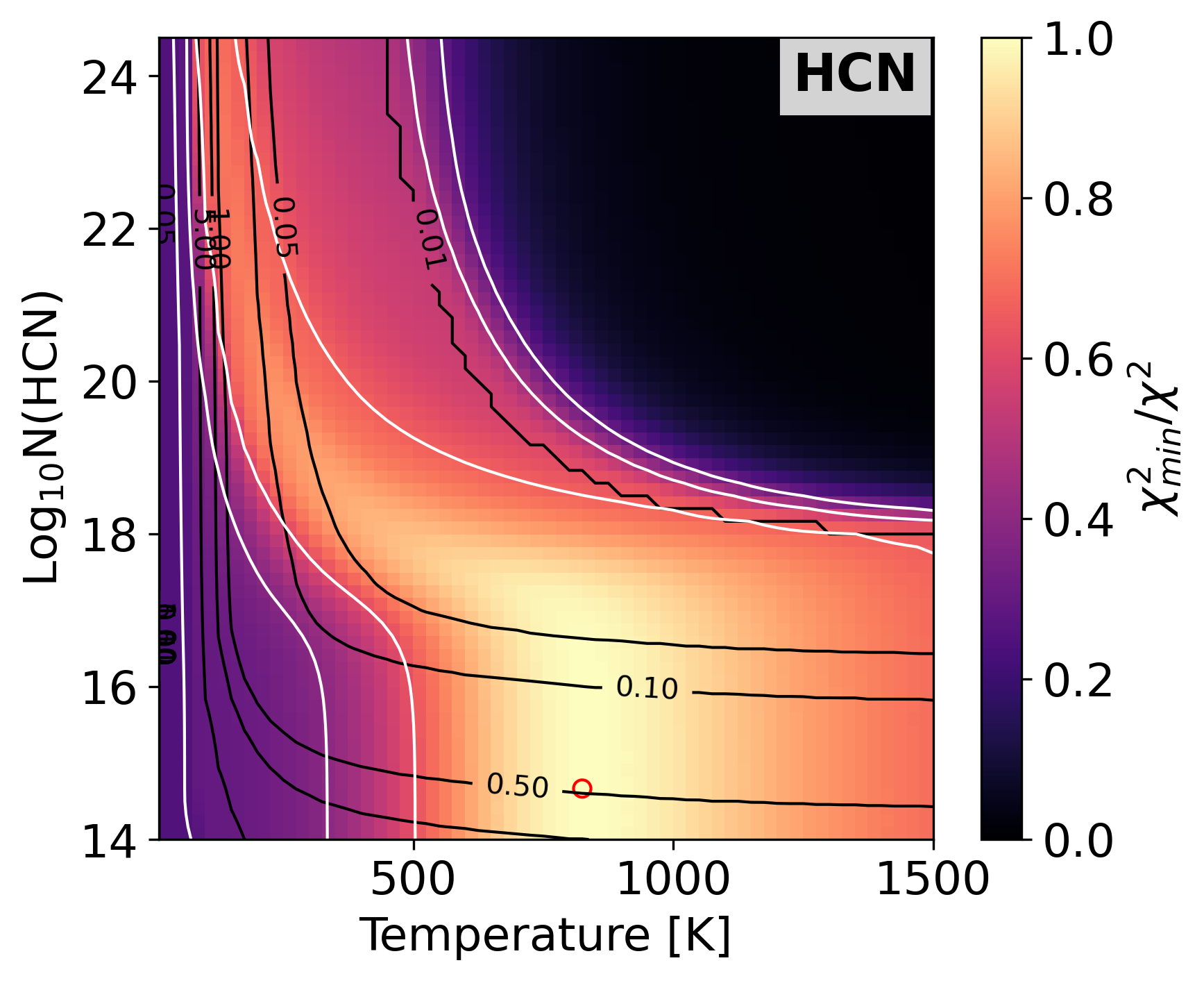}
&
\includegraphics[width=0.3\textwidth]{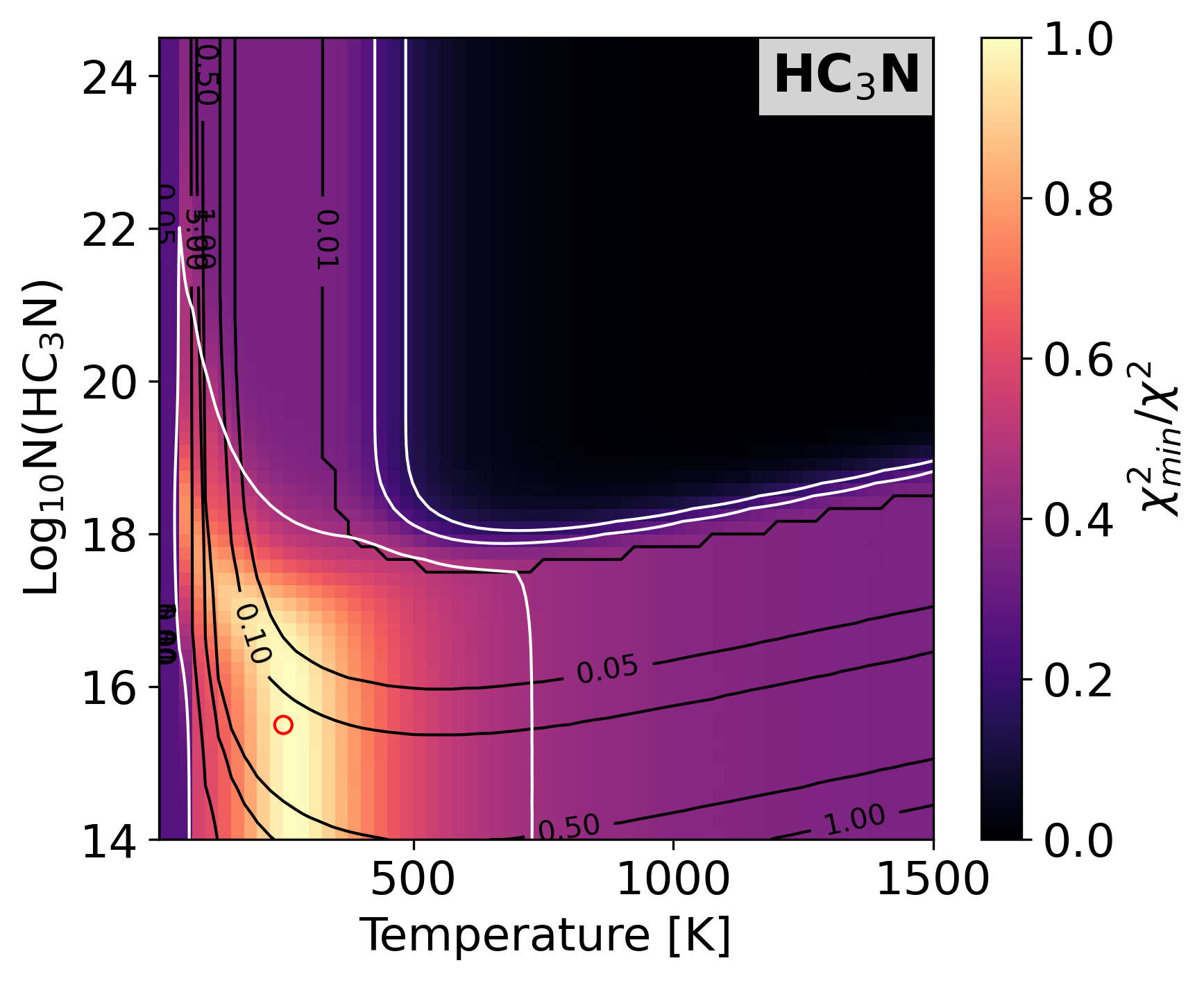}
&
\includegraphics[width=0.3\textwidth]{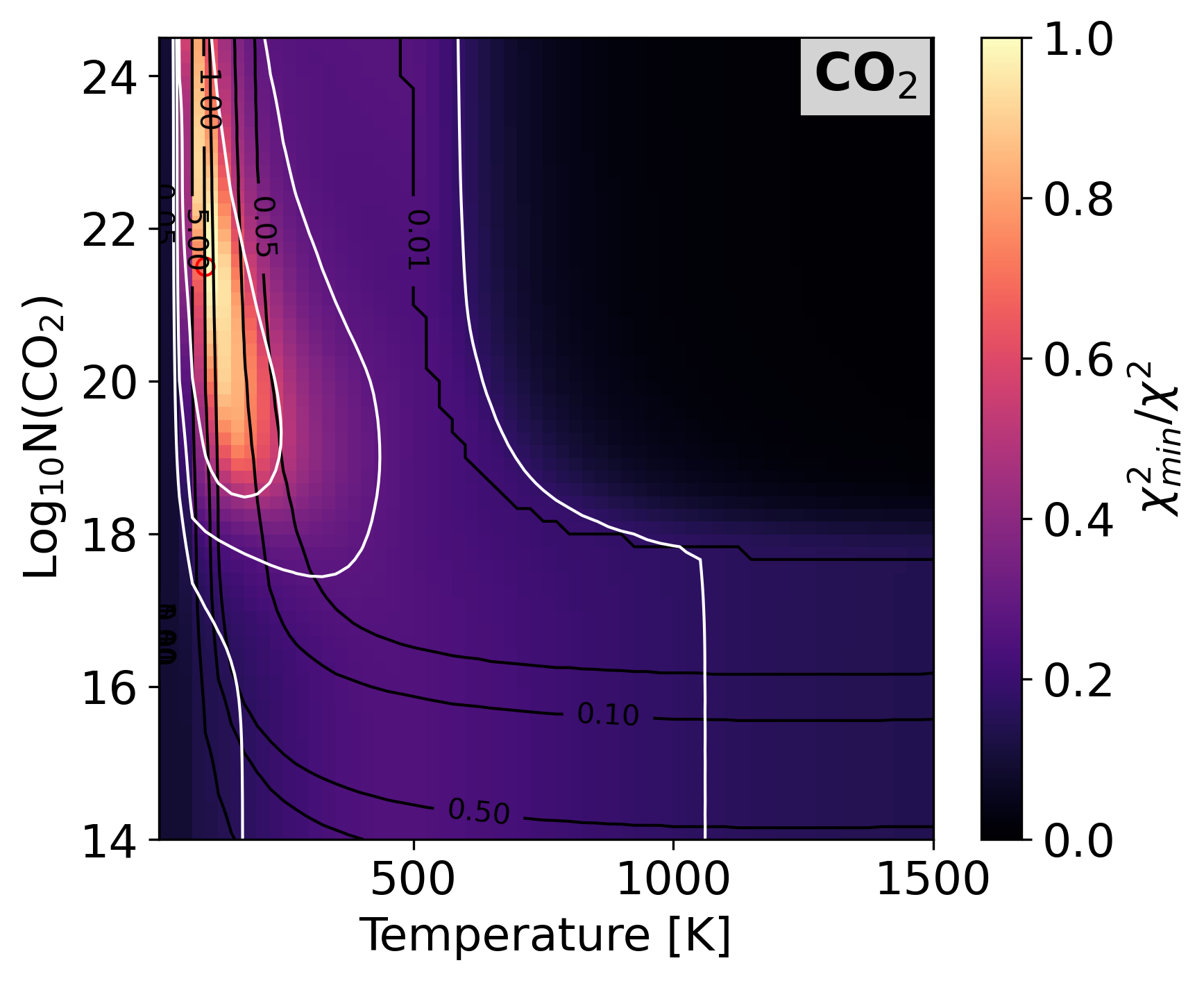}
\\
\includegraphics[width=0.3\textwidth]{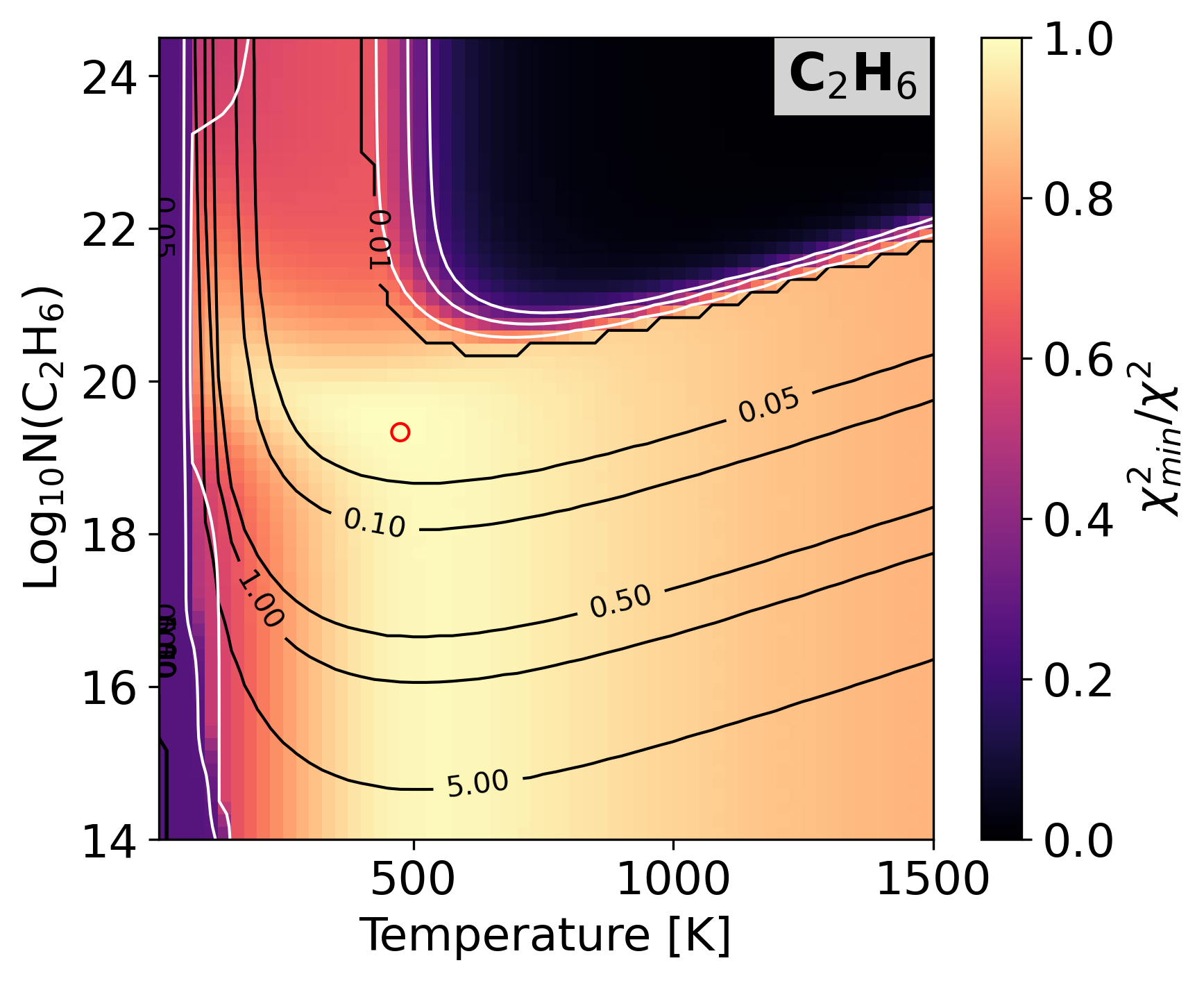}
&
\includegraphics[width=0.3\textwidth]{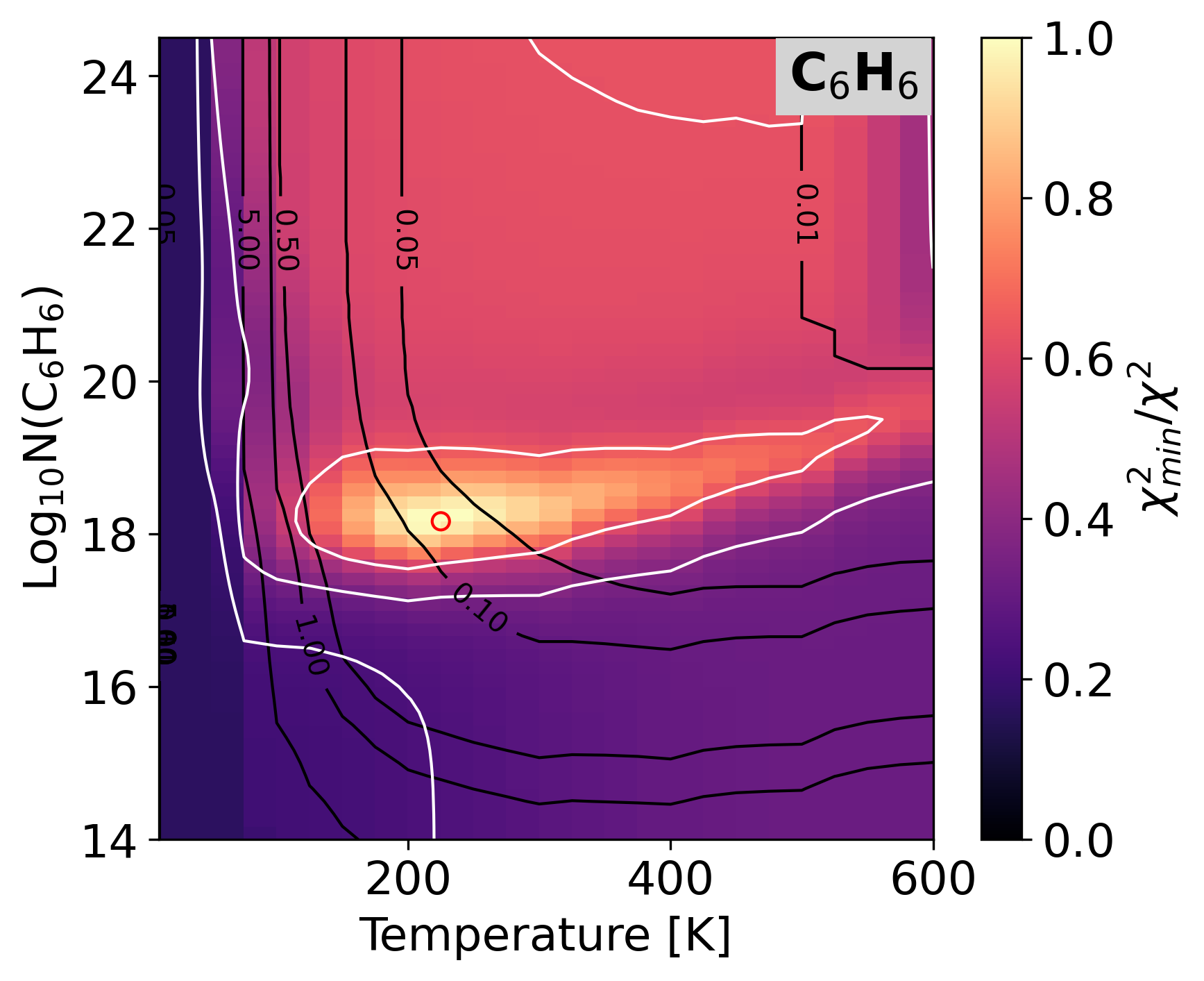}
\end{tabular}

\caption{The $\chi^2$ maps depicting the degeneracy in $T$ and $N$. The white contours represent $\chi^2_{min}$\,+\,2.3, $\chi^2_{min}$\,+\,6.2 and  $\chi^2_{min}$\,+\,11.8 confidence intervals. The black contours represent the emitting radii of 0.01, 0.05, 0.1, 0.5, 1, 5\,au. The red dot depicts the most representative set of conditions for the molecular emission.}
\label{chi2maps}
\end{figure*}

\section{Detailed analysis of \ce{CO2}}

\begin{figure*}[!h]
   \centering
    \includegraphics[width=\linewidth]{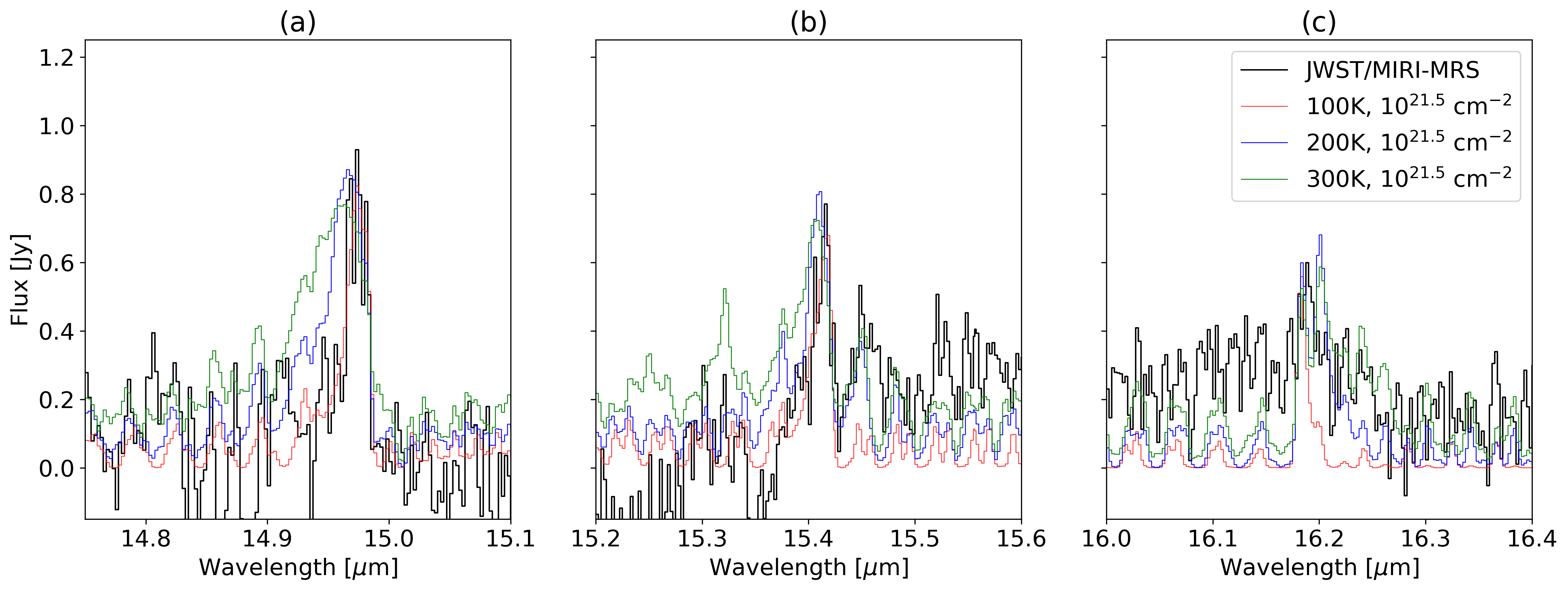}
   \caption{The \ce{CO2} emission with slab models at different $T$ at a constant $N$ where $R_{\mathrm{em}}$ is used as a scaling factor to match the flux levels. The $R_{\mathrm{em}}$ of 0.823\,au is used for the model in red, 0.069\,au and 0.028\,au for blue and green, respectively. Panel (a) shows the fundamental \ce{CO2} band, (b) shows the fundamental $^{13}$CO$_2$ band and (c) depicts the excited bending modes of \ce{CO2} and $^{13}$CO$_2$. The 100\,K (red) slab matches well with the JWST-MIRI/MRS data.}
              \label{CO2_blowup}%
    \end{figure*}

The JWST-MIRI/MRS spectrum of \ce{CO2}, $^{\rm{13}}$CO$_{\rm{2}}$ and the hot band of \ce{CO2} are shown in three different panels in Fig\,\ref{CO2_blowup}. The temperatures for \ce{CO2} can vary from 100 to 250\,K. The narrow width of the peak of \ce{CO2} drives the temperature solution towards 100\,K. The location of the Q-branch and the height of the hot band determines the column density $N$ \citep{Grant2023} and leads to values greater than 10$^{18.5}$\,cm$^{\rm{-2}}$; the emitting area is a scaling factor to best match the flux levels. 

\section{Chemistry with canonical and enhanced C/O ratio}
\label{Chemistry in model with canonical and enhanced C/O ratio}

\begin{table}[]
\caption{Parameters for the thermo-chemical disk model used in Sect.\,\ref{Hydrocarbon chemistry in disks} adopted from \cite{Greenwood2017}.}
\label{Parameter}
\resizebox{\linewidth}{!}{%
\begin{tabular}{lll} \hline
Quantity                            & Symbol                          & Values                             \\ \hline
\multicolumn{2}{l}{Stellar parameters}                                &                                    \\ \hline
stellar mass                        & $M_{\star}$                     & 0.12\,M$_{\sun}$                   \\
stellar luminosity                  & $L_{\star}$                     & 0.04\,L$_{\sun}$\tablefootmark{a}                    \\
effective temperature               & $T_{\mathrm{eff}}$              & 3060\,K                            \\
UV excess                           & $f_{\mathrm{UV}}$               & 0.01                               \\
UV powerlaw index                   & $p_{\mathrm{UV}}$               & 1                                  \\
strength of incident vertical UV    & $\chi^{\mathrm{ISM}}$            & 1                                  \\ 
cosmic ray \ce{H2} ionization rate  & $\zeta_{\rm CR}$                         & 1.3\,$\times$\,10$^{-17}$ s$^{-1}$ \\ \hline
\multicolumn{3}{l}{Disk parameters}                                                                        \\ \hline
minimum dust particle radius        & $a_\mathrm{min}$                & 0.05\,$\mu$m                        \\
maximum dust particle radius        & $a_\mathrm{max}$                & 3000\,$\mu$m                        \\
settling method                     & settle\textunderscore method    & \cite{Dubrulle1995}                  \\
settling parameter                  & $a_\mathrm{settle}$ or $\alpha$ & 10$^{-3}$                          \\
disk gas mass                       & $M_{\mathrm{disk}}$                      & 4.0\,$\times$\,10$^{-4}$\,M$_{\sun}$    \\
dust-to-gas ratio                   & dust-to-gas ratio               & 0.01                               \\
inner disk radius of the outer disk & $R_{\rm in}$                        & 0.035\,au                          \\
outer disk radius                   & $R_{\rm out}$                       & 30\,au                             \\
carbon-to-oxygen ratio              & C/O ratio                       & 0.45, 2.0                          \\
column density power index          & $\epsilon$                      & 1                                  \\
flaring index            & $\beta$                        & 1.15                                  \\
reference scale height    & $H_\mathrm{g}(100~{\rm au})$                           & 10\,au                             \\
extension                           & raduc                           & 1.15                               \\
maximum $\Sigma$ reduction          & reduc                           & 10$^{-7}$                          \\
distance                            & $d$                               & 192.2\,pc                          \\
inclination                         & $i$                               & 45\textdegree                          \\ 
grid size                           & radial\,$\times$\,vertical            & $300\,\times\,200$                       \\ \hline
\end{tabular}}
\tablefoot{These parameters are explained in detail in \cite{Woitke2009} and \cite{Woitke2023}. The two values of carbon-to-oxygen ratio corresponds to the canonical and the enhanced model, respectively. \tablefoottext{a}{The value for the luminosity differs from the reported value in \cite{Manara2017} as it is rescaled according to the new distance obtained from the Gaia 3 \citep{Galli2021} release.}}
\end{table}
\begin{figure*}[!h]
   \centering
   \includegraphics[width=\textwidth]{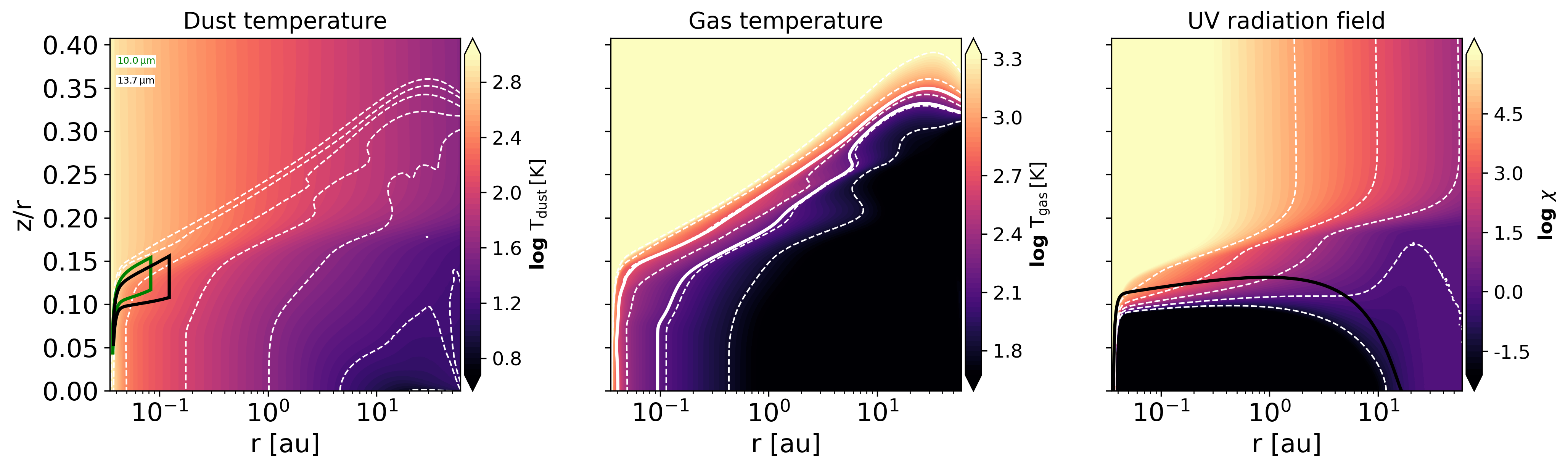}
   \caption{The dust temperature, gas temperature and UV radiation field calculated from radiative transfer for the canonical and enhanced C/O models. The green and black contour show the dust continuum at 10 and 13.7\,$\mu$m in first panel. The white contours in the second panel show the gas temperature corresponding to 140\,K and 475\,K. The black contour shows the A$_v$=1 mag in the last panel. The dotted white contours represent the values corresponding to the tick marks of the color scale.}
              \label{properties}%
\end{figure*}

\begin{figure}[!h]
   \centering
    \includegraphics[width=\linewidth]{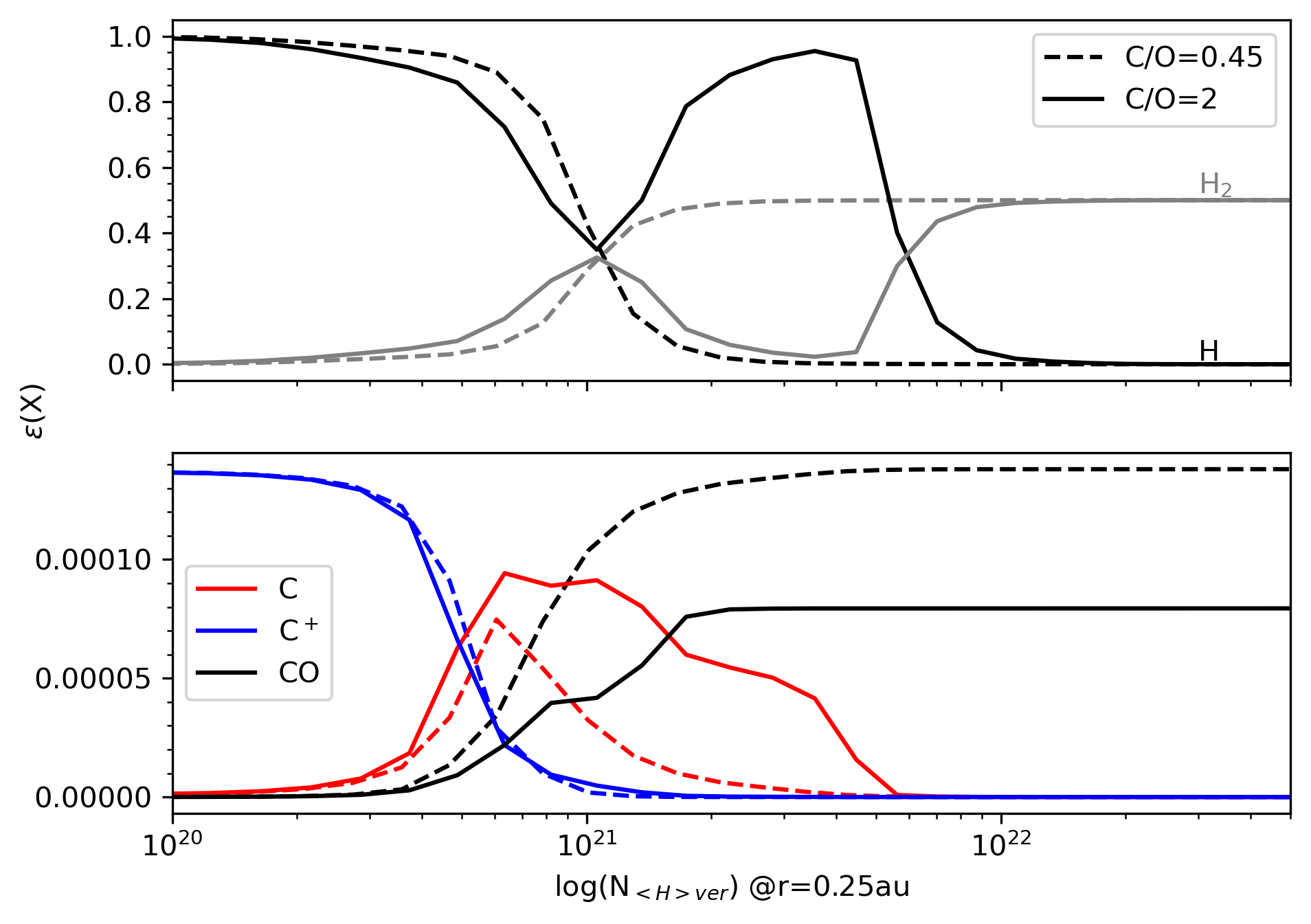}
   \caption{The H/\ce{H2} transition in the canonical (dashed lines) and enhanced C/O (solid lines) model at the radius of 0.25\,au are shown in the top panel. The bottom panel shows the \ce{C+}/C/CO transition layers in both models at the same radius.}
              \label{H_H2}%
    \end{figure}

Table\,\ref{Parameter} depicts the parameters used in the thermo-chemical disk models described in Sect.\,\ref{Hydrocarbon chemistry in disks}. The stellar luminosity is rescaled on the basis of the new distance reported in \cite{Galli2021}. Figure\,\ref{properties} shows the gas and dust temperature structure and the UV radiation field for both models. These are same in both the models. The canonical model has C/O\,=\,0.45 whereas the enhanced model has a C/O\,=\,2. Figure\,\ref{together} shows the contours with a value of 10\% of the maximum abundance of species in the canonical and enhanced C/O models. These maximum values of the abundance of species along with the maximum total column density and total column density at a radius of 0.1\,au in both the models are listed in Table \ref{max_col_density}. The maximum total column density is calculated by integrating all the molecules of a certain species from the surface to the midplane at every radii and the maximum of those are reported. Similarly, the total column density at 0.1\,au is calculated by integrating all the molecules of a certain species at the radius of 0.1\,au.

\begin{table*}[]
\caption{Maximum total column density ($N_{\rm{max}}$), radius at which $N_{\rm{max}}$ is reached, maximum abundance of species ($\epsilon_{\rm{max}}$) and the total column density at a radius of 0.1\,au ($N$) in models with canonical and enhanced C/O elemental ratio.}
\resizebox{\linewidth}{!}{%
\begin{tabular}{l|l|l|l|l|l|l|l|l} \hline
\multirow{2}{*}{Species} & \multicolumn{4}{c}{Canonical Model (C/O\,=\,0.45)} \vline    & \multicolumn{4}{c}{Enhanced Model (C/O\,=\,2)}        \\  [5pt] \cline{2-9}
                         & $N_{\rm{max}}\,\mathrm{(cm^{-2})}$ & r at $N_{\rm{max}}$ (au) & $\epsilon_{\rm max}$ & $N\,\mathrm{(cm^{-2})}$ at 0.1\,au & $N_{\rm{max}}\,\mathrm{(cm^{-2})}$ & r at $N_{\rm{max}}$ (au) & $\epsilon_{\rm max}$ & $N\,\mathrm{(cm^{-2})}$ at 0.1\,au\\ [5pt] \hline
\ce{CH3}                 & 1.3$\times 10^{16}$ & 3.7$\times$10$^{-2}$     & 9.2$\times$10$^{-8}$        & 2.0$\times 10^{12}$ & 4.2$\times 10^{16}$ & 3.7$\times$10$^{-2}$   & 1.8$\times$10$^{-7}$ & 2.8$\times 10^{14}$      \\
\ce{CH4}                 & 1.9$\times 10^{21}$ & 4.1$\times$10$^{-2}$    & 2.9$\times$10$^{-5}$        & 2.4$\times 10^{18}$ & 2.1$\times 10^{21}$ & 5.7$\times$10$^{-2}$   & 3.4$\times$10$^{-5}$ & 2.9$\times 10^{18}$    \\
\ce{C2H2}                & 6.3$\times 10^{20}$  & 4.2$\times$10$^{-2}$   & 7.1$\times$10$^{-6}$        & 9.9$\times 10^{13}$ & 7.2$\times 10^{20}$ & 4.5$\times$10$^{-2}$   & 7.1$\times$10$^{-6}$ & 2.1$\times 10^{17}$      \\
\ce{C2H4}                & 3.9$\times 10^{18}$ & 4.2$\times$10$^{-2}$    & 6.3$\times$10$^{-8}$        & 2.2$\times 10^{14}$ & 5.3$\times 10^{18}$ & 4.7$\times$10$^{-2}$   & 1.2$\times$10$^{-7}$ & 8.5$\times 10^{14}$      \\
\ce{C2H6}              & 6.1$\times 10^{21}$  & 4.2$\times$10$^{-2}$   & 3.0$\times$10$^{-5}$          & 2.1$\times 10^{13}$ & 4.8$\times 10^{21}$& 4.5$\times$10$^{-2}$    & 2.9$\times$10$^{-5}$ & 4.6$\times 10^{14}$       \\
\ce{C4H2}                & 8.1$\times 10^{12}$ & 4.0$\times$10$^{-2}$     & 7.9$\times$10$^{-10}$       & 1.8$\times 10^{7}$  & 2.4$\times 10^{15}$ & 3.6$\times$10$^{-2}$   & 1.4$\times$10$^{-8}$ & 2.0$\times 10^{13}$       \\
\ce{C3H4}              & 2.7$\times 10^{19}$ & 4.2$\times$10$^{-2}$    & 1.1$\times$10$^{-7}$          & 5.2$\times 10^{9}$  & 3.0$\times 10^{19}$ &  4.3$\times$10$^{-2}$  & 1.1$\times$10$^{-7}$ & 2.5$\times 10^{15}$      \\
\ce{C6H6}                & 1.0$\times 10^{19}$ & 4.2$\times$10$^{-2}$    & 6.9$\times$10$^{-8}$        & 7.0$\times 10^{5}$    & 1.8$\times 10^{19}$& 4.6$\times$10$^{-2}$   & 8.3$\times$10$^{-8}$ & 3.8$\times 10^{14}$       \\
\ce{HCN}                 & 2.5$\times 10^{21}$ & 4.2$\times$10$^{-2}$    & 3.3$\times$10$^{-5}$        & 1.7$\times 10^{16}$ &  3.5$\times 10^{21}$ & 4.6$\times$10$^{-2}$   & 7.2$\times$10$^{-5}$ & 1.8$\times 10^{19}$      \\ 
\ce{CO}                  & 3.5$\times 10^{22}$ & 5.2$\times$10$^{-2}$    & 1.4$\times$10$^{-4}$        & 3.6$\times 10^{20}$ & 2.3$\times 10^{22}$ & 5.2$\times$10$^{-2}$   & 7.9$\times$10$^{-5}$  & 1.7$\times 10^{20}$      \\
\ce{CO2}                 & 6.5$\times 10^{21}$  & 1.4$\times$10$^{-1}$   & 9.9$\times$10$^{-5}$        & 6.7$\times 10^{19}$ & 6.6$\times 10^{20}$ &  1.6$\times$10$^{-1}$  & 1.4$\times$10$^{-5}$  & 1.6$\times 10^{18}$     \\ 
\ce{H2O}                 & 7.0$\times 10^{22}$ & 4.2$\times$10$^{-2}$    & 3.0$\times 10^{-4}$         & 3.7$\times 10^{22}$ & 1.6$\times 10^{22}$ &  4.2$\times$10$^{-2}$  & 7.9$\times 10^{-5}$   & 1.4$\times 10^{19}$ \\ 
\ce{OH}               & 1.4$\times 10^{18}$   & 3.6$\times 10^{-2}$      & 3.7$\times 10^{-6}$ & 1.6$\times 10^{15}$ & 2.8$\times 10^{16}$ & 3.6$\times 10{-2}$ & 4.4$\times 10^{-7}$ & 2.6$\times 10^{14}$ \\\hline
\end{tabular}}
\tablefoot{The abundances are relative to the total hydrogen number density (n$_{<H>}$\,=\,$n$H+2$n$\ce{H2}, where n is number of particles).}
\label{max_col_density}
\end{table*}

Figure\,\ref{H_H2} shows the vertical H/\ce{H2} and the \ce{C+}/C/CO transitions at a distance of 0.25\,au. The H/\ce{H2} transition occurs deeper in the disk and the vertical range over which C remains abundant is extended in the enhanced C/O model. The H/H2 transition is mainly driven by the \ce{H2} formation on grains and photodissociation of \ce{H2} in the canonical model with C/O ratio of 0.45. The presence of unblocked carbon, H and \ce{H2} results in the formation of hydrocarbons in the disk atmosphere \citep{Kanwar2023}. \ce{H2} can then be destroyed through the abstraction of H  either by hydrocarbons or C itself in the enhanced C/O model. Thus, the H/\ce{H2} transition is pushed deeper into the disk to the point where all the left-over carbon is locked in CO.

In the following, we explain the analysis of the formation/destruction pathways for \ce{C2H2} and \ce{C6H6} in more detail. We study these in the models described in Sect.\,\ref{Thermo-chemical modelling}. The models with C/O ratios of 0.45 and 2 are referred to as canonical and enhanced model\textbf{s}, respectively. We analyse these pathways at a grid point that lies in the surface emitting layer of \ce{C2H2}. The characteristics describing the grid point are $T_\mathrm{gas}$\,=\,270\,K and $T_\mathrm{dust}$\,=\,230\,K, $n_\mathrm{<H>}$\,=\,6.4$\cdot \rm10^{10}cm^{-3}$, $A_\mathrm{V}^{ver}$\,=\,0.018, $A_\mathrm{V}^{rad}$\,= 1.1. 

The following set of reactions were dominant formation pathways for \ce{C2H2} in the model with the canonical C/O similar to the findings of \cite{Kanwar2023}:
\begin{equation}\label{R2}
    \ce{C2H3+} + \ce{e-} \rightarrow \ce{C2H2} + \ce{H}
\end{equation}
\begin{equation}\label{R3}
    \ce{O} + \ce{H2CCC} \rightarrow \ce{CO} + \ce{C2H2}
\end{equation}
\begin{equation}\label{R4}
    \ce{O} + \ce{C3H3+} \rightarrow \ce{HCO+} + \ce{C2H2}
\end{equation}
\begin{equation}\label{R5}
    \ce{C3H3+} + \ce{e-} \rightarrow \ce{CH} + \ce{C2H2}
\end{equation}
As evident from reactions\,\ref{R3} and \ref{R4}, the chemistry is driven by O. The contribution of reaction \ref{R3} in forming \ce{C2H2} decreased from $\sim$\,2\% in the canonical model to $\sim$\,0.01\% in the enhanced C/O ratio model. This is a result of depleting O to enhance the C/O ratio. The following reactions then become more important than reactions\,\ref{R3} and \ref{R4}:
\begin{equation}\label{R6}
    \ce{C} + \ce{C2H2+} \rightarrow \ce{C2H2} + \ce{C+}
\end{equation}
\begin{equation}
    \ce{C2H} + \ce{H} + \ce{M} \rightarrow \ce{C2H2} + \ce{M}
\end{equation}
These reactions were active in the canonical model but their rates increased as the elemental abundance of O is depleted in the enhanced C/O model. Hence, the dominant reaction pathways only shuffle in their relative importance when depleting O.

The destruction reactions for benzene in both models at the same grid point (see above) are:
\begin{equation}\label{B1}
    \ce{C6H6} + h\nu \rightarrow \ce{C2H2} + \ce{CH2CHCCH}
\end{equation}
\begin{equation}\label{B2}
    \ce{C+} + \ce{C6H6} \rightarrow \ce{C6H6+} + \ce{C}
\end{equation}
\begin{equation}\label{B3}
    \ce{C+} + \ce{C6H6} \rightarrow \ce{C5H3+} + \ce{C2H3}
\end{equation}
\begin{equation}\label{B4}
    \ce{C+} + \ce{C6H6} \rightarrow \ce{C7H5+} + \ce{H}
\end{equation}
Reaction\,\ref{B1} contributes $\sim$85\% to the destruction of \ce{C6H6} in the canonical C/O model which becomes $\sim$23\% in the enhanced C/O model. Reaction\,\ref{B2} is the next most dominant reaction contributing $\sim$10\% in the canonical C/O model which becomes $\sim$52\% in the model with enhanced C/O. This redistribution in the contribution is due to the increased abundance of \ce{C2H2} which promotes reaction~\ref{R6}, thus increasing \ce{C+}. The other way to form \ce{C+} is via UV photo-ionisation of C. The \ce{C2H2} abundance increase is due to the depletion of O and thus enhancement of the C/O ratio.

\begin{figure*}[!h]
   \centering
   \includegraphics[width=\linewidth,keepaspectratio]{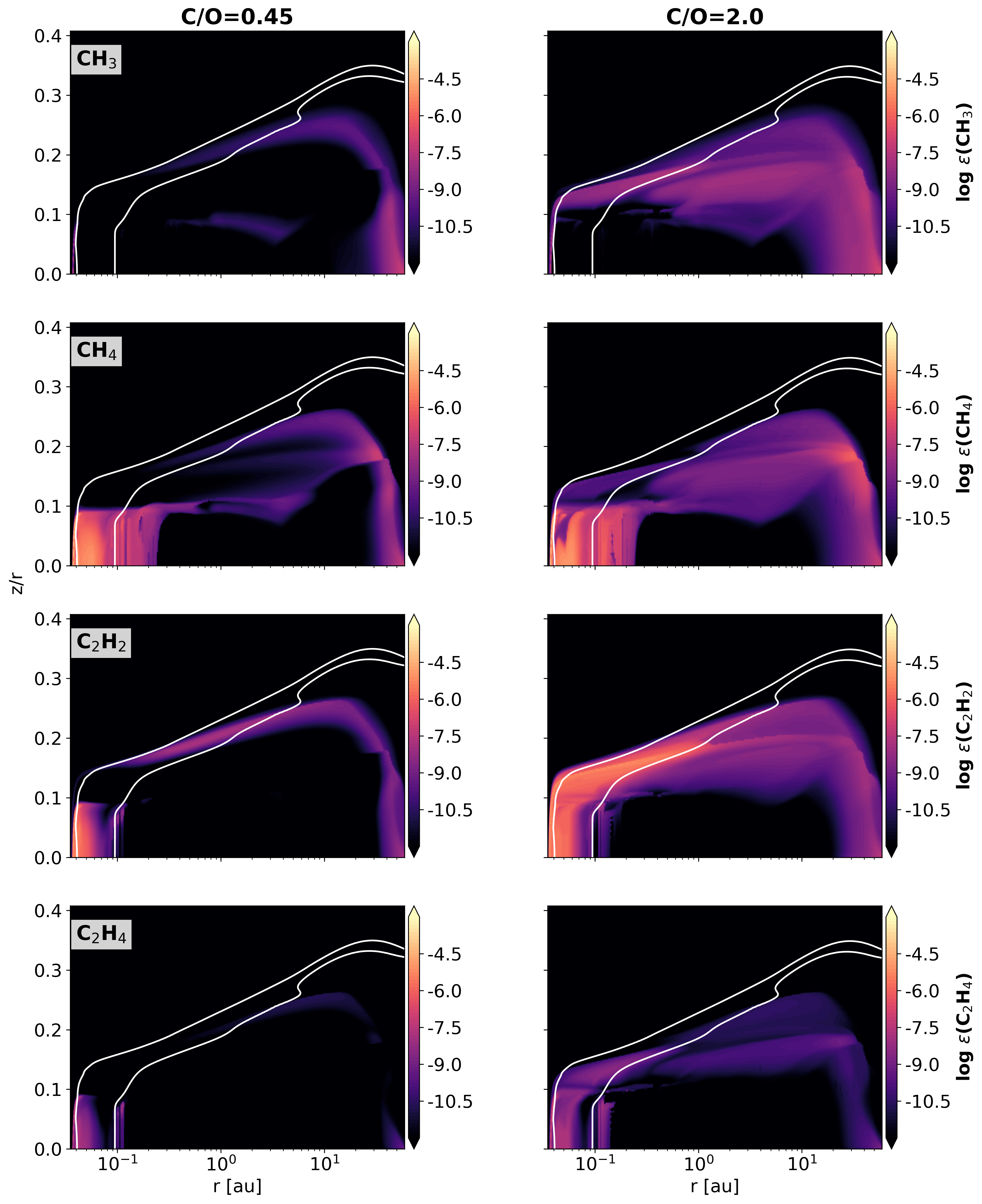}
   \caption{Abundances of various hydrocarbons in the canonical (0.45, left column) and enhanced (2.0, right column) C/O ratio models. The white contours corresponds to 140\,K and 475\,K gas temperature.}
              \label{Abundance_plots}%
\end{figure*}

\begin{figure*}[!h]
   \centering
   \includegraphics[width=\linewidth,keepaspectratio]{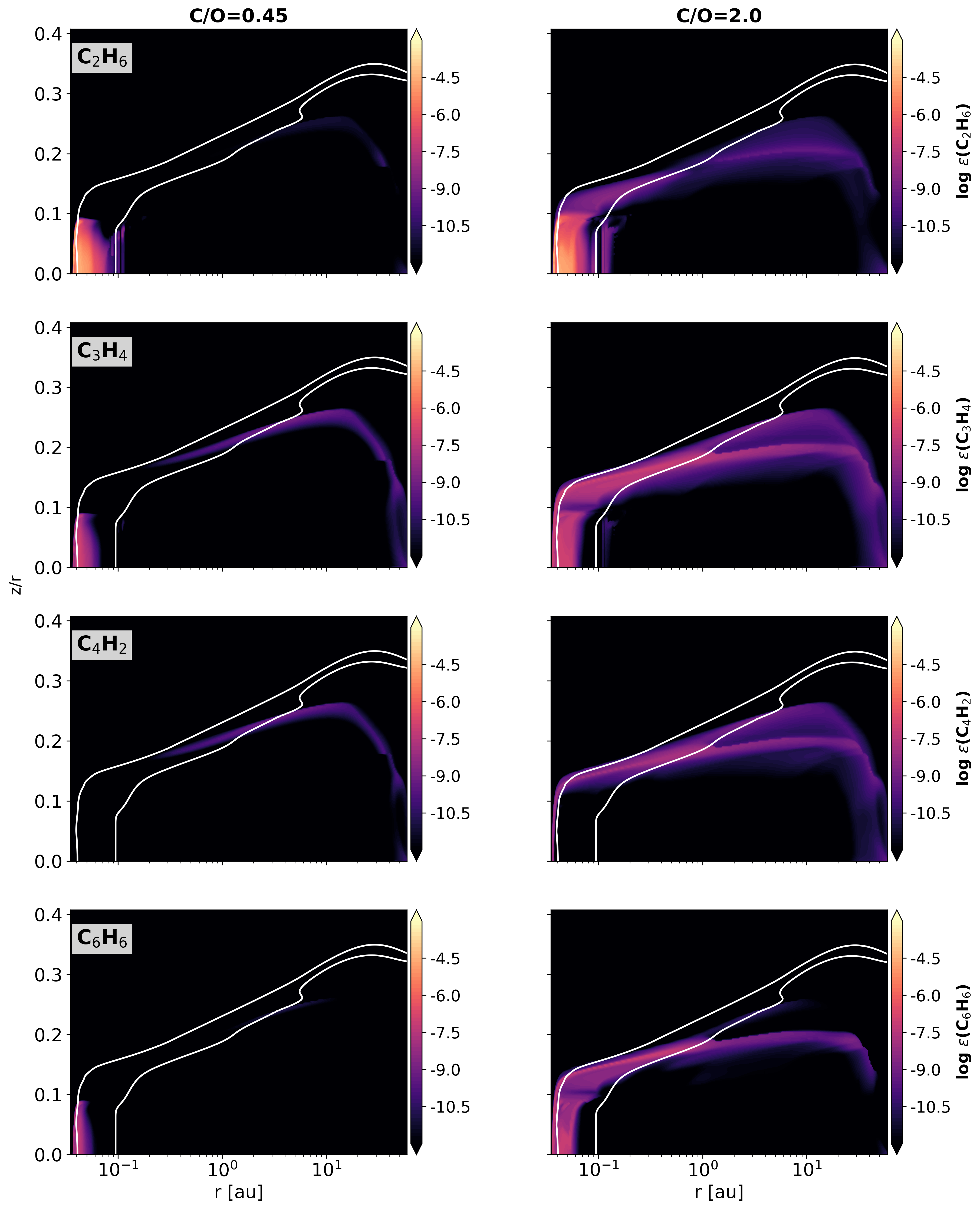}
   \caption{Abundances of various hydrocarbons in the canonical (0.45, left column) and enhanced (2.0, right column) C/O ratio models. The white contours corresponds to 140\,K and 475\,K gas temperature.}
              \label{Abundance_plots2}%
\end{figure*}

\begin{figure*}[!h]
   \centering
   \includegraphics[width=\linewidth,keepaspectratio]{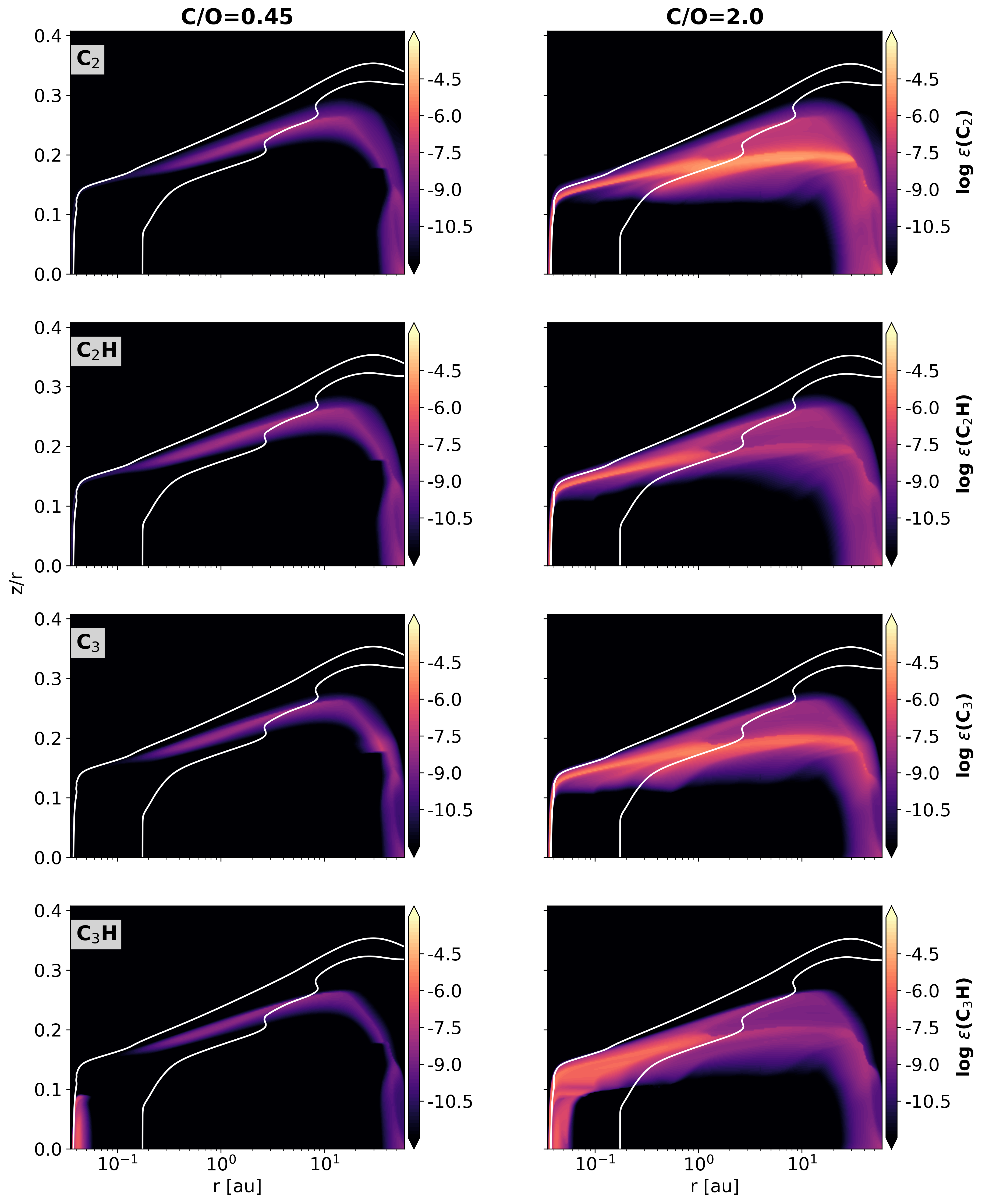}
   \caption{Abundances of various hydrocarbons in the canonical (0.45, left column) and enhanced (2.0, right column) C/O ratio models. The white contours corresponds to 140\,K and 475\,K gas temperature.}
              \label{Abundance_plots3}%
\end{figure*}

\begin{figure*}[!h]
   \centering
   \includegraphics[width=\linewidth,keepaspectratio]{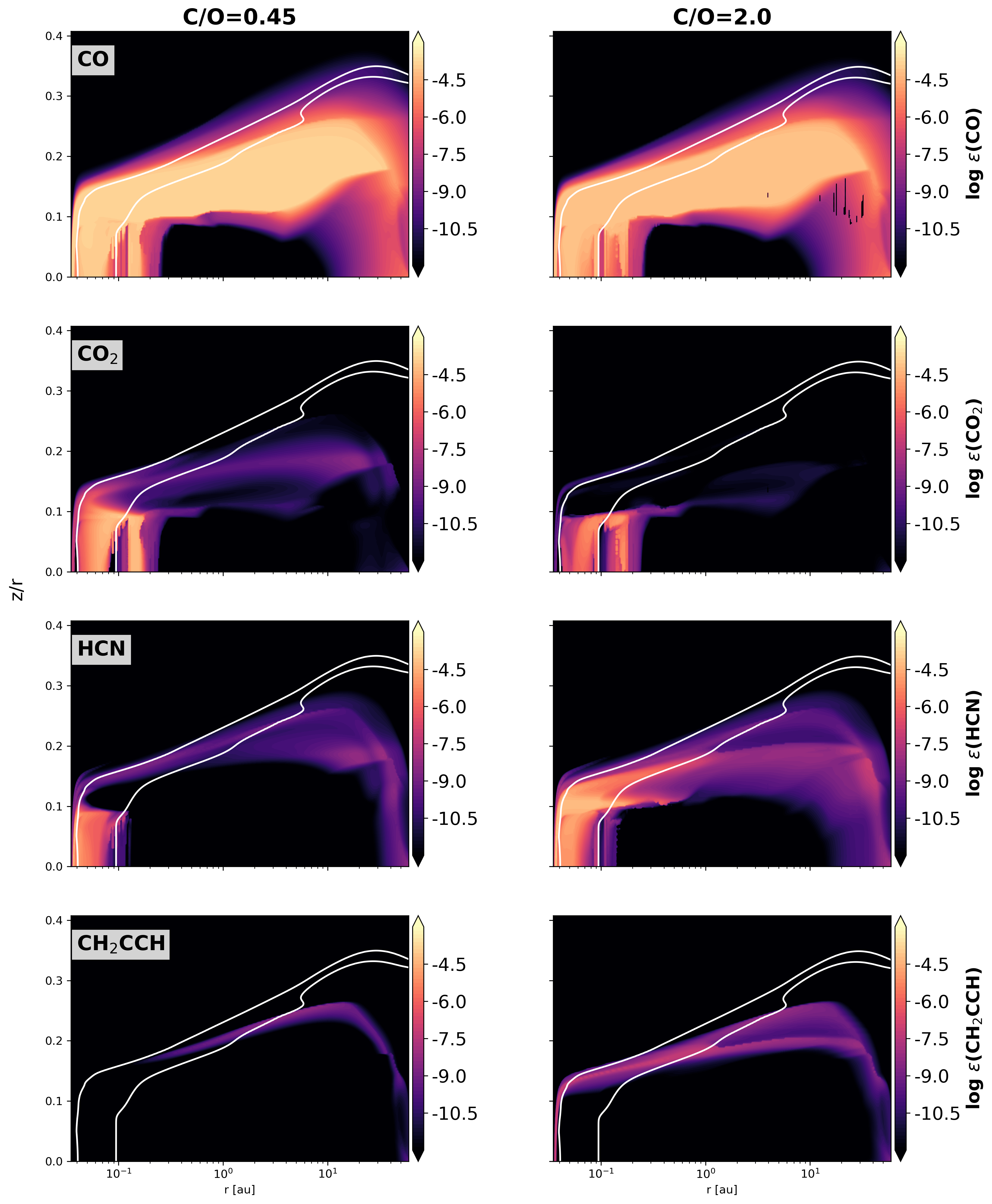}
   \caption{Abundances of various species in the canonical (0.45, left column) and enhanced (2.0, right column) C/O ratio models. The white contours corresponds to 140\,K and 475\,K gas temperature.}
              \label{Abundance_plots4}%
\end{figure*}

Figures\,\ref{Abundance_plots}, \ref{Abundance_plots2}, \ref{Abundance_plots3} and \ref{Abundance_plots4} show the 2D abundances of various species obtained from the thermo-chemical disk models. The abundances of the hydrocarbons increase in the surface layers when we change C/O from 0.45 (canonical) to 2.0 (enhanced). The blank regions in the abundance plot of CO for the enhanced model are due to the failure in solving the chemical reaction rate network at those locations. This might be because we do not model a fully self-consistent solution (the gas temperature structure is taken from the canonical model).


%
%

\end{document}